# Anomalous Hall magnetoresistance in a ferromagnet

Yumeng Yang[1], Ziyan Luo[1], Haijun Wu[2], Yanjun Xu[1], Run-Wei Li[3], Stephen J. Pennycook[2], Shufeng Zhang[4], and Yihong Wu[1,a)]

[1]*Department of Electrical and Computer Engineering, National University of Singapore, 4 Engineering Drive 3, Singapore 117583, Singapore*

[2]*Department of Materials Science & Engineering, National University of Singapore, Singapore 117575, Singapore*

[3]*Key Laboratory of Magnetic Materials and Devices, Ningbo Institute of Materials Technology and Engineering, Chinese Academy of Sciences, Ningbo 315201, People's Republic of China*

[4]*Department of Physics, University of Arizona, Tucson, Arizona 85721, USA*

The anomalous Hall effect, observed in conducting ferromagnets with broken time-reversal symmetry, offers the possibility to couple spin and orbital degrees of freedom of electrons in ferromagnets. In addition to charge, the anomalous Hall effect also leads to spin accumulation at the surfaces perpendicular to both the current and magnetization direction. Here we experimentally demonstrate that the spin accumulation, subsequent spin backflow, and spin-charge conversion can give rise to a different type of spin current related magnetoresistance, dubbed here as the anomalous Hall magnetoresistance, which has the same angular dependence as the recently discovered spin Hall magnetoresistance. The anomalous Hall magnetoresistance is observed in four types of samples: co-sputtered $(Fe_{1-x}Mn_x)_{0.6}Pt_{0.4}$, $Fe_{1-x}Mn_x$/Pt multilayer, $Fe_{1-x}Mn_x$ with $x = 0.17 – 0.65$ and Fe, and analyzed using the drift-diffusion model. Our results provide an alternative route to study charge-spin conversion in ferromagnets and to exploit it for potential spintronic applications.

a) Author to whom correspondence should be addressed: elewuyh@nus.edu.sg

Magnetoresistance (MR) in ferromagnetic (FM) materials and related heterostructures plays essential roles both in fundamental understanding of magnetism and electron transport in these structures and in various technological applications[1-4]. The most widely studied MR effects include anisotropic magnetoresistance (AMR), giant magnetoresistance (GMR) and tunnel magnetoresistance (TMR). These MR effects typically arise from spin-dependent transport of charge carriers either in the bulk or at the interfaces of these structures or the combination of both. Recently the discovery of several types of MR effects of different origins have triggered a renewed interest for spin-dependent MR; these include spin Hall magnetoresistance (SMR) in FM/heavy metal (HM) bilayers[5-10], Rashba-Edelstein magnetoresistance (REMR) in Bi/Ag/CoFeB[11], and Hanle magnetoresistance (HMR) in heavy metals[12]. One key aspect of these recently discovered MR effects is that they all originate from a two-step charge-spin conversion process, *i.e.*, in the first step charge current is converted to spin current through either the spin Hall effect (SHE)[13, 14] or the Rashba-Edelstein effect (REE)[15, 16], and in the second step part of the reflected spin current is converted back to charge current by the respective inverse effects. As inverse SHE (ISHE) always co-exists inside a material with SHE, the interplay of these two gives rise to an extra positive resistance contribution to a bulk conductor, which was first reported in the context of Hall effect in semiconductors[17]. In the proximity of surfaces/edges, part of the SHE generated spin current is cancelled out by the reflected spin current due to spin accumulation, resulting in a negative resistance contribution, as first derived by Dyakonov[18]. In the recently observed MR effects, the positive contribution does not play a role because it is insensitive to external field, while the negative contribution is modulated through controlling the amount of spin current reflection by either an adjacent magnetization (SMR and REMR) or an external magnetic field (HMR). Despite their small magnitude, these MRs are powerful tools to extract spin transport parameters, particularly spin-orbit torque (SOT) in FM/HM heterostructures[19-27], which has important applications in three-terminal[20], logic[28], and sensing[29, 30] devices. Unlike conventional MR effect, in all these MR effects, the FM plays a relatively less important role as both charge to spin and spin to charge conversions take place inside the HM layer. The FM only influences the conversion process indirectly through regulating the amount of spin current reflected back to FM/HM or FM/non-magnetic metal (NM) interfaces. From the application viewpoint, however, it will be of interest to investigate if a MR effect similar to SMR can be present in a FM alone, as this would allow additional flexibility in manipulating the charge-spin conversion process via controlling the magnetization of the FM directly.

Recently, anomalous Hall effect (AHE) in FM has attracted attention as an alternative mechanism for generating spin current or SOT in FM/NM multilayers. When a charge current $\mathbf{j_c}$ flows in an FM in the longitudinal direction, spin-up and spin-down electrons are deflected to opposite transverse directions via extrinsic mechanisms like skew scattering and side-jump or intrinsic mechanism related to band structure of the material[31]. Due to the asymmetry in density of states at the Fermi level and charge transport in FM, both transverse charge and spin accumulations will occur at boundaries of the sample at steady state. The former acts on the entire sample, generating the AHE voltage; while the latter leads to a backflow of spin current that only affects the vicinity of the sample boundary. Taniguchi *el al.*[32, 33] have predicted theoretically the presence of AHE-related SOT in FM/NM/FM trilayers and magnetoresistance in FM/NM bilayers. Very recently, several experimental attempts have been made to detect the AHE-induced spin current through either spin injection experiment in $Y_3Fe_5O_{12}$/Py heterostructure[34, 35] or characterization of SOT in FM/NM/FM sandwich structures[36, 37]. However, since all these experiments involve multiple layers, it is difficult to rule out completely contributions other than AHE to the predicted or observed MR or SOT. In this regard, here we report on a magnetoresistance induced by AHE and its inverse effect in a single FM layer, and refer it to as anomalous Hall magnetoresistance (AHMR). In order to observe the AHMR, one requires FM with a large AHE. Therefore, we focus on four types of FMs, *i.e.*, co-sputtered $(Fe_{1-x}Mn_x)_{0.6}Pt_{0.4}$, $Fe_{1-x}Mn_x$/Pt multilayers[38, 39], $Fe_{1-x}Mn_x$ with $x = 0.17 - 0.65$ and Fe. These materials are chosen because they allow to tune the saturation magnetization and thus the strength of AHE by simply adjusting the Mn composition (except for Fe). $Fe_{1-x}Mn_x$ itself can be tuned from ferromagnet to antiferromagnet by controlling the Mn composition. The inclusion of Pt further enhances the AHE in both the co-sputtered and multilayer samples. Magnetoresistance with SMR-like angular dependence is observed in all the four types of samples. We argue that the observed magnetoresistance is AHMR instead of SMR because all the samples behave as a single phase FM. Our argument is further substantiated by scaling analysis of the AHE and the relation between the MR and anomalous Hall angle. Based on the drift-diffusion formalism, we derive the analytical equations for MR in a single FM layer including AHE, and demonstrate that both the magnitude and thickness dependence of AHMR can be accounted for reasonably well using the analytical model.

## Results

**Angle dependent magnetoresistance.** As depicted in Figs. 1a and 1b, the applied charge current ($j_c$) in *x*-direction induces a transverse spin current ($j_s^t$) via AHE with the flow direction given by $\mathbf{m} \times \mathbf{j_c}$, where **m** is the magnetization direction. The simultaneous action of inverse AHE will convert a portion of $j_s^t$ back to charge current ($j_c'$) that has a direction opposite to the original one, thereby increasing the overall resistance of FM (positive contribution). For the case wherein **m**||**y**, with the comparable scale of film thickness and spin diffusion length, the backflow of spin current largely cancels $j_s^t$ and reduces the extra resistance (negative contribution). Whereas, when **m**||**z**, with the large lateral size, such cancellation is confined in the proximity of the sample edges only, and $j_s^t$ inside the sample remains nearly constant. To illustrate the difference in two cases, we illustrate in Figs. 1a and 1b the distribution of net $j_s^t$ in a colormap, the deeper the color the larger the net spin current. On the other hand, in the case of **m**||**x** (see Fig. 1c), there is no AHE. Therefore, when the magnetization rotates in the *yz*-plane, an angle dependent MR, *i.e.*, AHMR, appears and its dependence is expected to be the same as that of SMR. However, it should be noted that in the case of AHMR, both positive and negative contributions come from a single layer of FM material. The former is uniform throughout the sample, whereas the latter is dependent on the distribution of reflected spin current from the edges/surfaces, which is determined by the relative orientation of the magnetization with respect to the sample geometry and current direction.

To experimentally characterize the AHMR, we fabricated four types of samples (see Fig. 1d for illustration of different types of sample structures). Al the samples were deposited on $SiO_2$/Si substrates using sputtering and patterned into Hall bars by photolithography and liftoff techniques (see Methods for details). Through combined characterization of X-ray diffraction and high resolution scanning transmission electron microscopy, the samples were found to be polycrystalline with texture (see Supplementary Note 1). To minimize SHE from HM, all the samples were uncapped except for the $Fe_{1-x}Mn_x$/Pt multilayer samples which ends automatically with a Pt layer. In order to prevent the sample from oxidation, this final Pt layer was intentionally made slightly thicker than the rest of the Pt layers inside the stack. In general, samples with $x < 0.4 - 0.5$ (depending on the structure) exhibit global ferromagnetic behavior with an in-plane anisotropy at room temperature, and their temperature dependence of magnetization can be fitted well using a semi-empirical model[40] (see Supplementary Note 1). Angle dependent magnetoresistance (ADMR) measurements were performed by subjecting the sample to a rotational field of 30 kOe in the $zx$, $zy$, and $xy$ planes, and measuring the longitudinal resistance under a DC current (see illustrations in Fig. 2a). Shown in Figs. 2b – 2e are the typical ADMR results for each type of sample, i.e., $[Fe_{0.83}Mn_{0.17}(0.6)Pt(0.4)]_{10}$/Pt(1) multilayer, co-sputtered $(Fe_{0.71}Mn_{0.29})_{0.6}Pt_{0.4}(9)$, $Fe_{0.71}Mn_{0.29}(9)$ and Fe(9) (the number inside the parentheses indicates thickness in nanometer, and the repeating period of the multilayer sample is 10). The field dependent magnetoresistance (FDMR) results can be found in Supplementary Note 2. The angle $\theta_{ij}$ ($i, j = x, y, z$) denotes the angle between the rotating field and $i$-axis when the field rotates from $i$ to $j$-axis in the $ij$ plane, e.g., $\theta_{xy}$ refers to the angle with respect to $x$-axis when the field rotates from $x$- to $y$-axis. As can be seen from these results, MR($\theta_{zx}$) exhibits a $\sin^2\theta_{zx}$ symmetry which is expected for conventional anisotropic magnetoresistance (AMR) in FM, whereas MR($\theta_{zy}$) is similar to SMR in FM/HM bilayer with $-\sin^2\theta_{zy}$ symmetry (see solid lines in Figs. 2b – 2e for fitting). Furthermore, the magnitude of MR($\theta_{xy}$) is the sum of the magnitude of MR($\theta_{zx}$) and MR($\theta_{zy}$) (the small difference may be due to slightly different saturation state in the three rotation directions). These features are in good agreement with the SMR observed in metallic FM/HM bilayers[41,42]. However, what is striking is that the same type of MR behavior was observed in all four types of samples despite their significantly different sample structures. In fact, in the case of Fe$_{0.71}$Mn$_{0.29}$(9) and Fe(9), there is even no heavy metal element involved at all. These results suggest that the MR($\theta_{zy}$) observed in these samples must have an origin different from the SMR. We shall mention that similar MR($\theta_{zy}$) was observed before in Fe/MgO[43] and MgO/Fe/MgO or SiO$_2$/Fe/SiO$_2$[44], but no unified explanation was given. Before ending this section, it is worth pointing out that there is a small deviation from the $\sin^2\theta_{zx}$ or $-\sin^2\theta_{zy}$ dependence in the fitting data shown in Figs. 2b – 2e, and the deviation increases with the saturation magnetization. Numerical simulation by using the experimentally derived demagnetizing field in z-direction ($H_d$) confirms that this is caused by the slight deviation of the magnetization direction from the external field direction when it is rotating in the zx- or zy-plane, though the latter (30 kOe in this case) is much higher than $H_d$[45]. Nevertheless, this deviation only alters the shape of the ADMR curves, which does not affect the magnitude of the extracted MR ratio (see Supplementary Note 3).

**Correlation of MR($\theta_{zy}$) and AHE.** To examine whether the observed MR($\theta_{zy}$) originates from AHE, we conducted scaling analysis by measuring MR($\theta_{zy}$) in samples with fixed thickness but different AHE strength. Specifically, we varied the AHE strength by adjusting the Mn composition in FeMn/Pt, FeMnPt and FeMn samples, and Pt composition in Fe$_{1-x}$Pt$_x$ samples (see Methods for more details). Although systematic studies have been performed on all these samples, here we only focus on the co-sputtered FeMnPt alloy samples in the main text as it is more representative as compared to the other three types of samples. The discussion on Fe$_{1-x}$Mn$_x$/Pt multilayer and Fe$_{1-x}$Mn$_x$ samples can be found in the Supplementary Note 4. The thickness of all samples is fixed at 9 nm. Fig. 3a shows the ADMR when the magnetization rotates in the $zy$-plane, *i.e.*, MR($\theta_{zy}$), for co-sputtered (Fe$_{1-x}$Mn$_x$)$_{0.6}$Pt$_{0.4}$ samples with $x$ = 0.22 – 0.65. As can be seen from the figure, the ADMR exhibits the same angle-dependence in the entire Mn composition range, though its magnitude decreases monotonically with increasing $x$. In addition, we investigated how longitudinal resistivity ($\rho_{xx}$), anomalous Hall resistivity ($\rho_{xy}^{AH}$) and $M_s$ vary with $x$ and the results are summarized in Figs. 3b and 3c, respectively. The $\rho_{xy}^{AH}$ values in Fig. 3c are obtained from the raw Hall resistivity $\rho_{xy}$ after subtracting out the contribution from ordinary Hall effect (see Supplementary Note 5). The anomalous Hall resistivity of FM is known to be proportional to the saturation magnetization in the same material system[31], *i.e.*, $\rho_{xy}^{AH} = R_s M_s$, where $R_s$ is the anomalous Hall coefficient and $M_s$ is the saturation magnetization. It should be noted that among different material systems, such relation may not apply. The AHE in FM can originate from either intrinsic or extrinsic mechanisms. The former arises from the electronic band structure or Berry phase, whereas the latter is due to scattering of charges by impurity or defects with large spin-orbit coupling, via either skew-scattering or side-jump mechanism[31]. The extrinsic mechanism is presumably dominant in the present case considering the structure of the samples. In this case, it has been established previously that $R_s \propto \rho_{xx}$ for skew-scattering and $R_s \propto \rho_{xx}^2$ for side-jump[31], provided that $M_s$ is constant. Since in the present case $M_s$ is changing with Mn composition, in Fig. 3d, we plot $\rho_{xy}^{AH}/M_s$ as a function of $\rho_{xx}$. As can be seen from the fitting, $\rho_{xy}^{AH}/M_s$ scales almost linearly with $\rho_{xx}$ for $x < 0.6$, which suggests that skew scattering is indeed dominant in this case. The deviation for samples with $x > 0.6$ (see inset of Fig. 3d), is due to the weakening of FM order as evident from the drastic drop of $M_s$ in Fig. 3c. In analogy to spin Hall angle ($\theta_{SH}$) of HM, we can define an anomalous Hall angle as $\theta_{AH} = \frac{\sigma_{xy}^{AH}}{\sigma_{xx}}$ or $\frac{\rho_{xy}^{AH}}{\rho_{xx}}$, with $\sigma_{xy}$ and $\sigma_{xy}^{AH}$ the anomalous Hall and longitudinal conductivity of FM, respectively. By using the experimentally determined $\theta_{AH}$ values, in Fig. 3e, we plot the amplitude of MR($\theta_{zy}$), *i.e.*, $\Delta R/R_{xx}$, as a function of $\theta_{AH}$, for (Fe$_{1-x}$Mn$_x$)$_{0.6}$Pt$_{0.4}$ (squares). It is apparent that the curve is non-linear, suggesting that the observed ADMR is a second order process of AHE, different from conventional AMR. It is worth noting that the maximum $\theta_{AH}$ value of 0.03 is comparable to the lower end of spin Hall angles reported for Pt (which itself is scattered over a large range)[46], but is two times as large as $\theta_{AH}$ of NiFe/Pt bilayer[47]. In the case of NiFe/Pt, spin-orbit coupling at the interface has been cited as the cause for enhanced $\theta_{AH}$. In the present case, however, the enhancement of $\theta_{AH}$ in $(Fe_{1-x}Mn_x)_{0.6}Pt_{0.4}$ is presumably due to Pt atoms uniformly distributed in the alloy films. In addition to $(Fe_{1-x}Mn_x)_{0.6}Pt_{0.4}$, we also show the results for $Fe_{1-x}Mn_x$, $Fe_{1-x}Pt_x$ and Fe in Fig. 3e (represented by different symbols). We will discuss these results shortly after presenting the analytical model. Besides these samples, some other common ferromagnetic and antiferromagnetic materials including Co, NiFe, and $Ir_{0.2}Mn_{0.8}$ were also examined, but they all exhibit a $\theta_{AH}$ at least one order of magnitude smaller, and therefore either very small or different MR($\theta_{zy}$) behavior was observed (see Supplementary Note 6). This is expected because the size of $\rho_{xy}^{AH}$ in transition metals typically follows the order: Fe >> Co > Ni[48-50]. As discussed in the Supplementary Note 7, field misalignment is not able to account for the magnitude of the measured MR($\theta_{zy}$) curves. Apart from the field misalignment, another possible source for the MR($\theta_{zy}$) observed is the geometric size effect (GSE) related AMR. However, if this is indeed the case, one would expect a same temperature dependence of MR($\theta_{zy}$) and MR($\theta_{zx}$). But, as shown in Supplementary Fig. 12, we observed a different temperature dependence for MR($\theta_{zy}$) and MR($\theta_{zx}$) in the FeMnPt and Fe samples, but same temperature dependence in the Py control sample which has a much smaller AHE. In view of these results, both field misalignment and GSE related AMR can be ruled out as the origin of the observed MR($\theta_{zy}$).

**Derivation of AHMR.** In order to have a quantitative understanding of the results shown in Fig. 3e, we derive the analytical equation for MR in a single FM layer by including the AHE and its inverse effect (see Supplementary Note 8). As discussed, a transverse spin current $j_s^t$ is generated in the direction of $\mathbf{m} \times \mathbf{j_c}$ when the charge current $\mathbf{j_c}$ flows in an FM, where $\mathbf{m}$ is the magnetization direction. In the case of bulk, $j_s^t$ is uniformly distributed inside the sample, which gives an extra resistance due to the additional opposite charge current induced by the inverse AHE. However, due to spin accumulation and backflow of spin current from the boundary, the situation changes when the sample has a finite dimension in the $j_s^t$ flowing direction. As derived in Supplementary Note 8, this will lead to a magnetoresistance that is given in the general form of

$$\rho_{xx} = \rho_0 \left(1 + A m_x^2 + (\theta_{AH}/\beta)^2 \left[m_z^2 + \left(1 - \frac{2l_s}{d}\tanh\left(\frac{d}{2l_s}\right)\right) m_y^2\right]\right) \qquad (1)$$

where $d$ and $l_s$ are the thickness and spin diffusion length of FM, respectively, $\beta$ is the polarization for longitudinal conductivity, $\theta_{AH}$ is the anomalous Hall angle, and $A$ is the AMR ratio. Apparently, Eq. (1) contains both AMR and AHMR contributions. In order to have an anatomical view of the spin-charge conversion process that leads to the AHMR, we plot the normalized spin accumulation $\mu_s/\mu_s(0)$, spin current $j_s^t/j_c$, and charge current $(j_{cx} - j_c)/j_c$ in Figs. 4a – 4c for **m**||**y**, and in Figs. 4d – 4f for **m**||**z**, respectively. Insets of Figs. 4d – 4f are the distributions near the sample edges. For the case of **m**||**y**, $j_s^t$ flows in z-direction with the spin polarization in y-direction, and it accumulates at the top and bottom surfaces. Under the boundary condition $j_s^t = 0$ at both surfaces, the spatial distribution of spin accumulation ($\mu_s$), transverse spin current ($j_s^t$) and longitudinal charge current ($j_{cx}$) in z-direction are given by

$$\mu_s(z) = \frac{2el_s j_c}{\sigma_{xx}} \frac{\theta_{AH}}{\beta} \frac{\cosh\left(\frac{z}{l_s}\right) - \cosh\left(\frac{z-d}{l_s}\right)}{\sinh(d/l_s)} \qquad (2)$$

$$j_s^t(z) = \frac{j_c \theta_{AH}}{\beta} \left[1 - \frac{\sinh\left(\frac{z}{l_s}\right) - \sinh\left(\frac{z-d}{l_s}\right)}{\sinh(d/l_s)}\right] \qquad (3)$$

$$j_{cx}(z) = j_c - j_c \left(\frac{\theta_{AH}}{\beta}\right)^2 \left[1 - \frac{\sinh\left(\frac{z}{l_s}\right) - \sinh\left(\frac{z-d}{l_s}\right)}{\sinh(d/l_s)}\right] \qquad (4)$$

where $j_c$ is the original applied charge current in x-direction, $\sigma_{xx}$ is the conductivity, $e$ is the electron charge, and the rest of parameters are already defined as above. The second term in the brackets of Eqs.

(3) and (4) is resulted from the backflow of spin current induced by the spin accumulation described by Eq. (1). The same set of equations applies to the case when **m**||**z** except that $d$ is replaced by the sample width ($w$), and the spatial distribution is along $y$-direction. In the calculations, we have used $d = 10$ nm, $w = 100$ μm, $\beta = 0.5$, $l_s = 3$ nm, and $\theta_{AH} = 0.03$. The values used for polarization and spin diffusion length are within the range of those reported in FMs[51]. As shown in Fig. 4b for **m**||**y**, where the dashed lines are added as a reference to show the case when AHE is absent in the sample, due to the comparable scale of $d$ and $l_s$, the backflow spin current cancels $j_s^t$ largely throughout the sample (see Fig. 1a for illustration). In contrast, in the case of **m**||**z**, the cancellation is mainly confined in the vicinity of the two side edges: $j_s^t$ in the remaining region remains almost intact because $w \gg l_s$ (see Fig. 4e and Fig. 1b). It is this difference in the cancellation of $j_s^t$ that leads to the different degree of charge current correction, which consequently results in the different resistance for **m**||**y** and **m**||**z**: the origin of AHMR. On the other hand, the AHE is absent when **m**||**x**, and therefore no transverse spin current / spin accumulation nor redistribution of charge current occurs in this case. Although the AHE does not come into play when **m**||**x**, the conventional AMR still exits and gives rise to an increase in resistance, which is revealed by the 2$^{nd}$ term in Eq. (1). Therefore, the AHMR, which is given by the 3$^{rd}$ term of Eq. (1), has the same angular dependence as SMR, in qualitative agreement with the experimental data shown in Figs. 2b – 2d and Fig. 3a. Notably, the size of AHMR ratio, given by $(\frac{\theta_{AH}}{\beta})^2 \frac{2l_s}{d} \tanh(\frac{d}{2l_s})$, exhibits a quadratic relationship with $\theta_{AH}$, which is in good agreement with the ADMR data shown in Fig. 3e. These results affirm our argument that the ADMR in $zy$-plane is caused by the AHE and its inverse in the FM layer, *i.e.*, the AHMR. It is apparent from Eq. (1) that, in addition to the experimentally determined anomalous Hall angle $\theta_{AH}$, the magnitude of AHMR is also directly dependent on $\beta$ and $l_s$, which are not available experimentally for most of the materials under investigation except for Fe. And in fact even for Fe, the reported values are scattered in a large range depending on the techniques used, film thickness, techniques used to prepare the film, *etc*. Therefore, in Fig. 3e, we first plot the range of AHMR calculated by using $\beta = 0.4 - 0.6$ and $l_s = 2 - 5\ nm$ (shadowed region). The lower and upper boundary denoted the minimum and maximum values obtained by $\beta = 0.6$, $l_s = 2\ nm$ and $\beta = 0.4$, $l_s = 5\ nm$, respectively. The range for $\beta$ and $l_s$ are chosen to cover most of the ferromagnets. Therefore, besides $(Fe_{1-x}Mn_x)_{0.6}Pt_{0.4}$, we also added the results for $Fe_{1-x}Mn_x$, $Fe_{1-x}Pt_x$, and Fe in the same figure (all have a thickness of 9 nm). The anomalous Hall angle were varied by changing either the Mn or Pt composition (except for Fe). Despite the variation in composition, the AHMR for all these Fe-based films indeed show a quadratic dependence on $\theta_{AH}$, as manifested in the dotted line, which is the fitting result for $(Fe_{1-x}Mn_x)_{0.6}Pt_{0.4}$ obtained by using $\beta = 0.55$ and $l_s = 3.5\ nm$. As we will discuss shortly, similar range of values can also fit the thickness-dependence of AHMR as predicted by Eq. (1).

**Thickness dependence of AHMR.** Fig. 5a shows the experimentally observed thickness dependence of AHMR for $(Fe_{0.71}Mn_{0.29})_{0.6}Pt_{0.4}$ with $d = 2 - 15$ nm. Instead of a monotonic decrease of AHMR with increasing $d$ as predicted by the theoretical model, the experimental value increases sharply at small thickness, peaks at around $d = 3$ nm, and then decreases slowly as $d$ increases. There are two possible reasons that cause the deviation from theoretical model at small thickness: one is the sharp increase of resistivity due to surface scattering and the other is the decrease of magnetization due to finite size effect. When the thickness of a thin film becomes smaller than or comparable to the electron mean free path, its resistivity scales with the thickness as $\rho_{xx} = \rho_{xx0}\left[1 + (1-p)\frac{3l_f}{8(d-d_0)}\right]$, here $\rho_{xx0}$ is the bulk resistivity, $p$ is the specular reflectivity, $d_0$ is the roughness, and $l_f$ is the electron mean free path[52]. For surface with finite roughness or small $p$, the resistivity will increase sharply when $d < l_f$ or $d_0$. Fig. 5b shows $\rho_{xx}$ and $M_s$ as a function of $d$. As expected, $\rho_{xx}$ increases, whereas $M_s$ decreases sharply at small thickness. It is interesting to note that $M_s$ starts to decrease at a larger thickness than $\rho_{xx}$, understandably from the difference in length scale that governs the resistivity and magnetization of thin films. More discussion on the effect of thin film roughness can be found in Supplementary Note 9. In Fig. 3, we found that the relation $\theta_{xy}^{AH} \propto M_s$ holds for most of the Mn composition range for FeMnPt, suggesting that the AHE is dominated by skew scattering[31]. For comparison, we plot, in Fig. 5c, $\rho_{xy}^{AH}/M_s$ as a function of $\rho_{xx}$ for samples with different thicknesses. A nearly perfect linear relation is obtained when $d > 3$ nm. However, at $d < 3$ nm (see inset of Fig. 5c), a sublinear relation appears, suggesting gradual weakening of AHE in this region. This is understandable because, in this region, surface scattering dominates the electrical transport; but compared to bulk scattering, surface scattering may not be an efficient mechanism for AHE since it is mostly spin-independent. Surface effect was not taken into account when deriving Eq. (1); therefore, strictly speaking, it does not apply to the case when the film thickness becomes comparable to or smaller than the spin diffusion length which is usually larger than the mean free path. Fig. 5d shows the experimentally determined anomalous Hall angle $\theta_{AH}$ as a function of $d$. It is almost a constant above $d = 5$ nm, but decreases rapidly below this thickness. If we take the average value of $\theta_{AH} = 0.026$ for $d = 5 - 15$ nm, we are able to fit the MR-dependence on thickness well using Eq. (1) for $d > 5$ nm. The fitting curve is shown in Fig. 5a as dotted line, in which we have used $\beta = 0.58$, $l_s = 4.5$ nm. These values are in the same range as those that are used for the fitting in Fig. 3 (dotted line) though they are not exactly the same (presumably due to thickness-dependent surface effect). The deviation at very small thickness corresponds to the region where surface scattering becomes dominant, leading to a sharp increase of resistivity and decrease of $\beta$ [53]. In the same region, $M_s$ decreases rapidly as well due to the decrease in Curie temperature. As the decrease of both $M_s$ and $\beta$ leads to a more rapidly decrease of $\theta_{AH}$ as compared to $\beta$ itself, the AHMR diminishes rapidly at small thickness. Therefore, in addition to the quadratic dependence on $\theta_{AH}$ presented in Fig. 3e, the thickness dependence of AHMR shown in Fig. 5a also strongly supports the AHE origin of MR($\theta_{zy}$). To further substantiate this argument, we have fabricated another series of $(Fe_{0.71}Mn_{0.29})_{0.6}Pt_{0.4}$ sample as well as a series of Fe samples in the thickness range of 5 – 20 nm. Again, a monotonically decreasing MR($\theta_{zy}$) ratio is obtained in the entire thickness range of $d = 5 - 20$ nm for both series of samples (see Supplementary Note 10 for more details). All these results combined confirm the reproducibility of the experimental results and validity of the AHMR scenario presented in this work.

## Discussion

As discussed above, the AHMR has been observed in co-sputtered $(Fe_{1-x}Mn_x)_{0.6}Pt_{0.4}$, $Fe_{1-x}Mn_x$/Pt multilayer, $Fe_{1-x}Mn_x$ with $x = 0.17 – 0.65$ and Fe. The resistivity and magnetization of these samples were varied systematically using both the Mn composition and layer thickness, which in turn allows us to use scaling analysis to examine the AHE origin of the observed MR. Both the magnitude and thickness dependence of the MR (> spin diffusion length) are in reasonable agreement with those calculated from an analytical model based on the drift-diffusion formalism. The FeMnPt based materials were chosen because they exhibit a relatively large AHE, and importantly the size of AHE can be tuned by adjusting the chemical composition. However, on the other hand, these materials are relatively new and their magnetic and electrical properties are less understood across the composition range. It will be of importance to confirm if the AHMR is also present in other others whose magnetic properties have been thoroughly investigated and well understood. One possible approach is to tune the AHE of Py by adjusting the Fe composition and see if there is any correlation between AHE and MR($\theta_{zy}$), though the weak AHE may pose a challenge in interpreting the experimental data. Another candidate for investigating the AHMR is the class of materials with giant AHE reported recently[54, 55]. In addition to electrical measurement, it would also be of interest to probe the AHE-generated spin current directly using magneto-optical technique and correlate it with the MR data. We believe that the results described in this work demonstrate the importance of AHE as an alternate tool for studying spin-charge interconversion in magnetic materials and its potential in spintronic applications.

## Methods

**Sample preparation.** All samples were deposited on $SiO_2$(300 nm)/Si substrates using DC magnetron sputtering with a base and working pressure of $2\times10^{-8}$ Torr and $3\times10^{-3}$ Torr, respectively. $Fe_{1-x}Mn_x$ (or $Fe_{1-x}Pt_x$) films were prepared by co-sputtering of $Fe_{0.8}Mn_{0.2}$ and Mn targets (or Fe and Pt targets). $[Fe_{1-x}Mn_x(0.6)/Pt(0.4)]_{10}$ multilayer samples were prepared by sequential deposition of $Fe_{1-x}Mn_x$ and Pt layers in a repeated manner, while $(Fe_{1-x}Mn_x)_{0.6}Pt_{0.4}$ samples were deposited by co-sputtering of $Fe_{0.8}Mn_{0.2}$, Mn, and Pt. The chemical composition of all the samples was determined by X-ray photoelectron spectroscopy (XPS). Standard photolithography and liftoff techniques were used to fabricate the Hall bar, which consists of a mesa 1.1 mm long and 100 μm wide, with four 50 μm wide and 200 μm long protrusions placed at sides of the bar as voltage probes. A Microtech laserwriter system with a 405 nm laser was employed to directly expose the substrates after coating the negative photoresist Microposit S1805. After exposure, the substrates were then soaked in developer MF319 to form the Hall bar pattern. After deposition using sputtering, the photoresist was removed by the mixture of PG remover and acetone, and the metallic patterns are left on the substrates.

**Characterization.** Structural properties of the samples were characterized using a Rigaku X-ray diffraction (XRD) system with Cu Kα radiation. X-ray photoelectron spectroscopy (XPS) was performed on PHI Quantera II XPS Scanning Microprobe from Ulvac-PHI with a beam spot size of 50 μm. In addition, high resolution scanning transmission electron microscopy (STEM, a JEOL ARM200F) was employed to directly image the multilayer samples. Magnetic properties were characterized using a Quantum Design vibrating sample magnetometer (VSM) with the samples cut into a size of 4 mm × 3 mm. The resolution of the system is better than $6\times10^{-7}$ emu. The electrical measurements were also performed using the same Quantum Design system at a bias current of 100 μA.

**Data availability.** The data that support the findings of this study are available from the corresponding author on request.

## Acknowledgements

Y.H.W. would like to acknowledge support by the Singapore National Research Foundation under its Competitive Research Programme (Grant No. NRF-CRP10-2012-03) and Ministry of Education, Singapore under its Tier 2 Grant (Grant No. MOE2013-T2-2-096). S.Z. is supported by U.S. National Science Foundation under Grant No. ECCS-1708180. S.J.P. is grateful to the National University of Singapore for funding. We thank Shengyuan A. Yang of Singapore University of Technology and Design for fruitful discussions. Y.H.W. is a member of the Singapore Spintronics Consortium (SG-SPIN).

## Author contributions

Y.H.W. designed and supervised the project. Y.Y. and Y.H.W. designed the experiments. Y.Y. and Y.X. fabricated the samples. Y.Y. performed the electrical and magnetic measurements. Z.L. performed the X-ray characterizations. H.W. and S.J.P. performed the high resolution scanning transmission electron microscopy characterizations and analyzed the data. S.Z., Y.H.W. and Y.Y. developed the AHMR model and analyzed the data. All authors discussed the results. Y.Y. and Y.H.W. wrote the manuscript and all the authors contributed to the final version of manuscript.

**Competing interests:** The Authors declare no Competing Financial or Non-Financial Interests.

**FIGURE CAPTIONS**

**Fig. 1** Illustration of AHE and inverse AHE in thin FM films and sample structures used in this study. AHE and inverse AHE in different magnetization configurations: **a**, **m**||**y**; **b**, **m**||**z**; and **c**, **m**||**x**. **d**, Schematics of three types of samples including co-sputtered $(Fe_{1-x}Mn_x)_{0.6}Pt_{0.4}$, $Fe_{1-x}Mn_x$/Pt multilayer, $Fe_{1-x}Mn_x$ with $x = 0.17 - 0.65$ and Fe. The colormap in **a** – **c** is the distribution of net $j_s^t$: the deeper the color the larger the net spin current.

**Fig. 2** ADMR measurement geometry and typical results. **a**, Schematic of measurement geometry. **b-d**, ADMR measurement results: **b**, $[Fe_{0.83}Mn_{0.17}(0.6)Pt(0.4)]_{10}/Pt(1)$ multilayer; **c**, co-sputtered $(Fe_{0.71}Mn_{0.29})_{0.6}Pt_{0.4}(9)$; **d**, $Fe_{0.71}Mn_{0.29}(9)$; and **e,** Fe. Solid lines in **b** - **e** are fittings based on the angle dependence.

**Fig. 3** Correlation of MR($\theta_{zy}$) and AHE. **a**, MR($\theta_{zy}$) results for co-sputtered $(Fe_{1-x}Mn_x)_{0.6}Pt_{0.4}$ with $x = 0.22 - 0.65$. **b** - **c,** Mn composition dependence of $\rho_{xx}$, $\rho_{xy}^{AH}$ and $M_s$, respectively. **d**, $\rho_{xy}^{AH}/M_s$ as a function of $\rho_{xx}$. **e,** MR ratio as a function of $\theta_{AH}$ for a variety of samples including Fe, $Fe_{1-x}Pt_x$, $(Fe_{1-x}Mn_x)_{0.6}Pt_{0.4}$ and $Fe_{1-x}Mn_x$. Solid lines in **a** are fittings based on the angle dependence. Inset of **d** shows the full range of $\rho_{xy}^{AH}/M_s$ vs. $\rho_{xx}$ plot, and solid line in **d** serves as a guide for the eye. The shadowed area in **e** is the calculated AHMR ratio range using different combinations of $\beta$: 0.4 - 0.6 and $l_s$: 2 - 5 nm, and dotted line in **e** is the fitting results using $\beta = 0.55$ and $l_s = 3.5$ nm. The error bars in **d** are from the linear fitting to determine $\rho_{xy}^{AH}$ from $\rho_{xy}$.

**Fig. 4** Modulation of the spin accumulation, transverse spin current, longitudinal charge current by AHE in FM. **a** – **c**, normalized distributions along $z$-direction for the cases of **m**||**y**: **a**, spin accumulation

$\mu_s/\mu_s(0)$; **b**, spin current $j_s^t/j_c$; and **c**, charge current $(j_{cx} - j_c)/j_c$. **d** – **f**, normalized distributions along $y$-direction for the cases of **m**||**z**: **d**, spin accumulation $\mu_s/\mu_s(0)$; **e**, spin current $j_s^t/j_c$; and **f**, charge current $(j_{cx} - j_c)/j_c$. Dashed lines in **a** – **f** are the reference when no AHE is present in the sample. Insets of **d** – **f** are the distributions near the vicinity of the edges.

**Fig. 5** Thickness dependence of AHMR and AHE in co-sputtered $(Fe_{0.71}Mn_{0.29})_{0.6}Pt_{0.4}$ samples. **a**, AHMR ratio for $(Fe_{0.71}Mn_{0.29})_{0.6}Pt_{0.4}$ with $d = 2 - 15$ nm, and the fitting using Eq. (4) with fixed $\theta_{AH}$ for $d = 5 - 15$ nm (dotted line). **b**, Thickness dependence of $M_s$ and $\rho_{xx}$. **c**, $\rho_{xy}^{AH}/M_s$ as a function of $\rho_{xx}$ in the linear range. **d,** Thickness dependence of $\theta_{AH}$. Inset of **c** shows the full range of $\rho_{xy}^{AH}/M_s$ vs. $\rho_{xx}$ plot, and solid line in **c** serves as a guide for the eye.

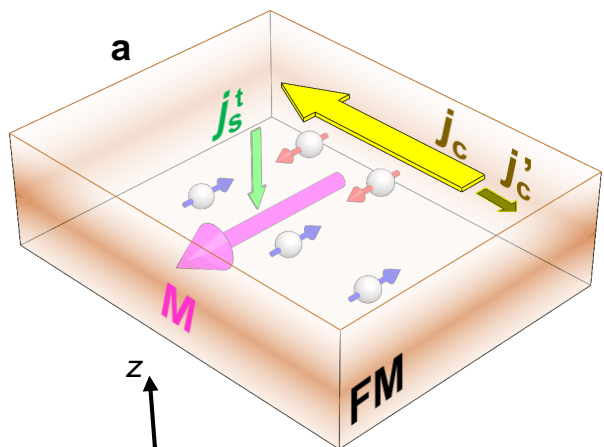 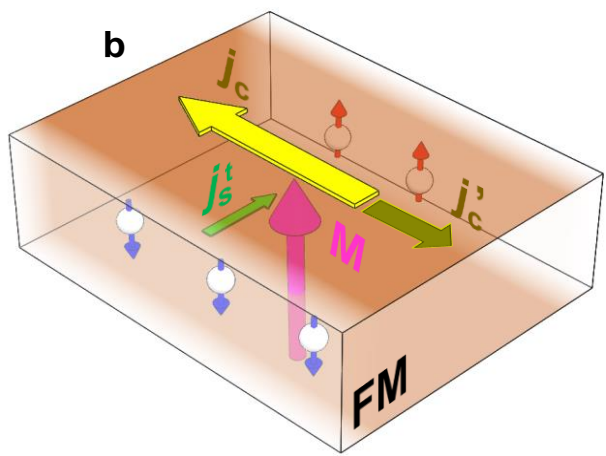 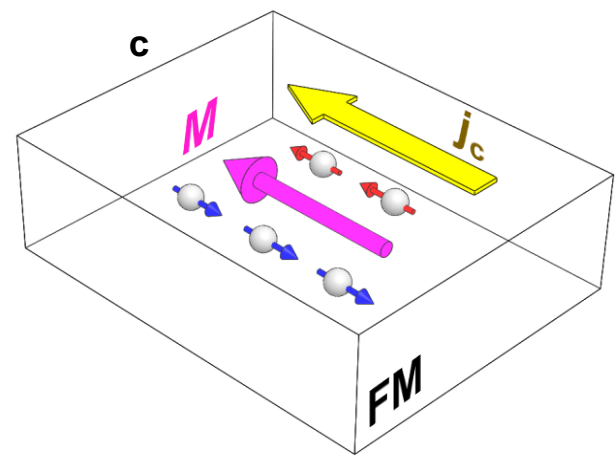
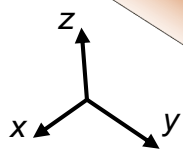
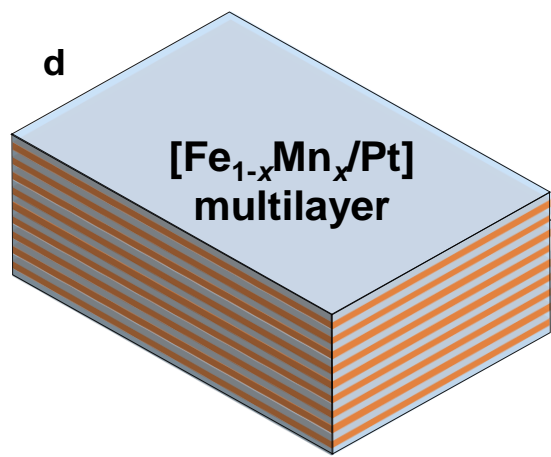 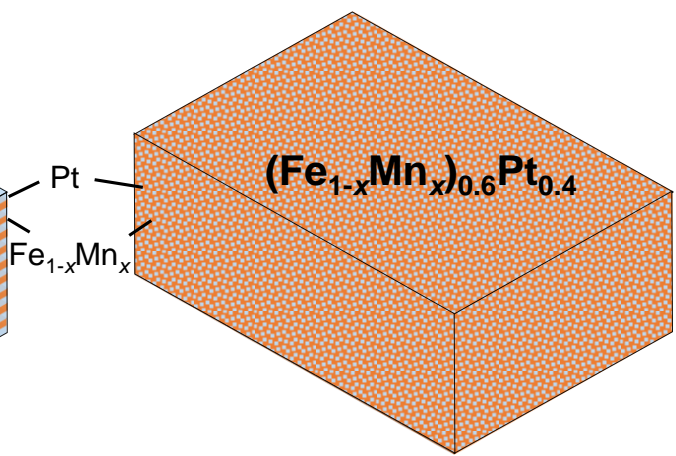 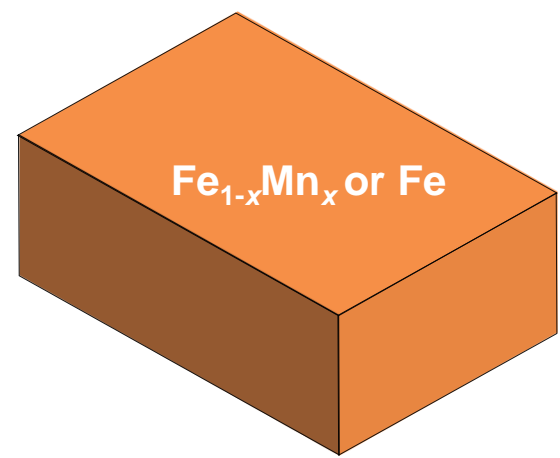

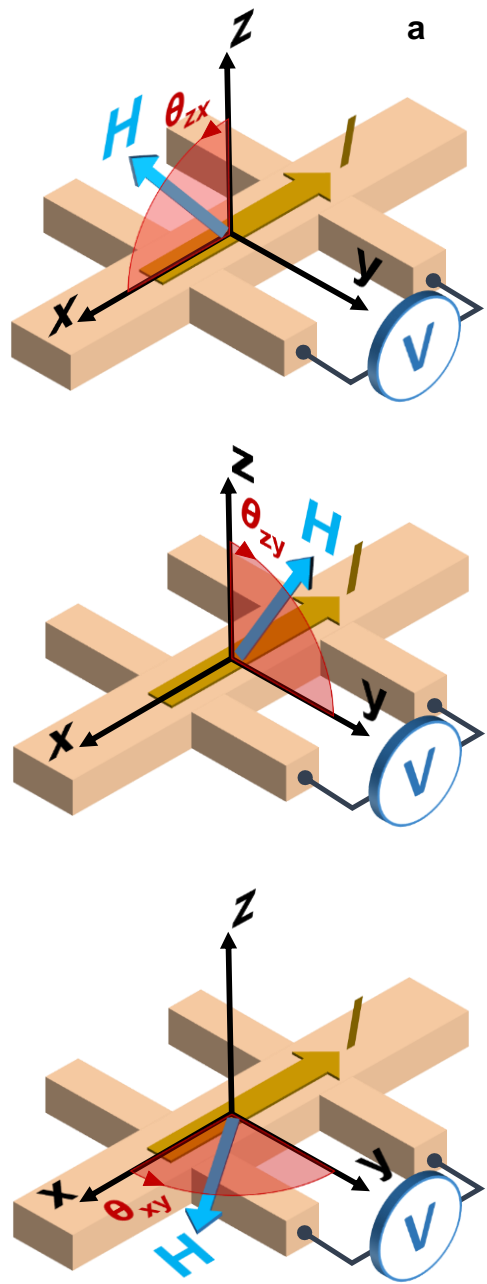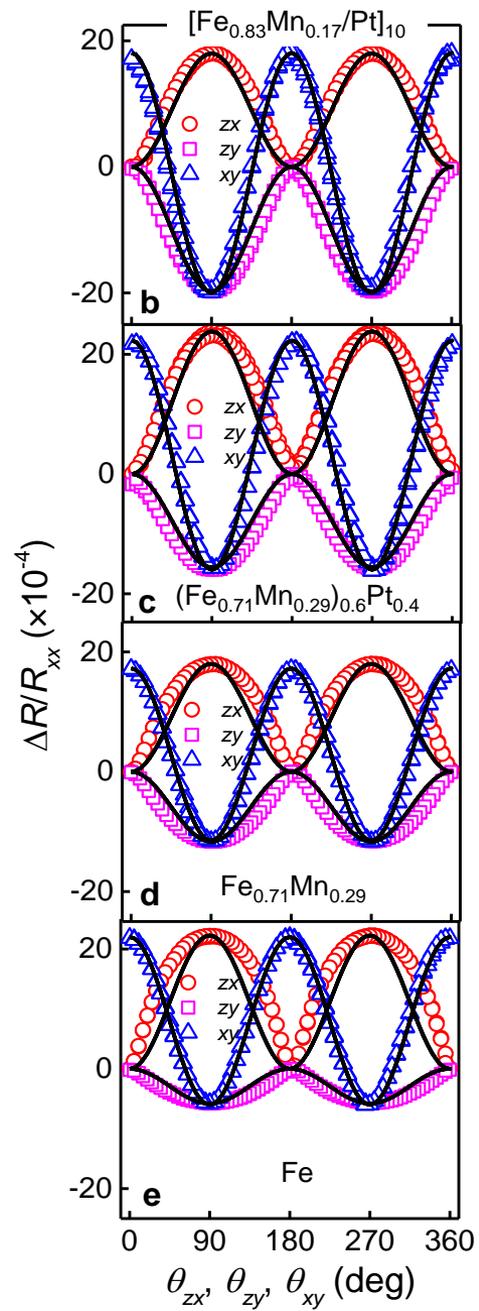

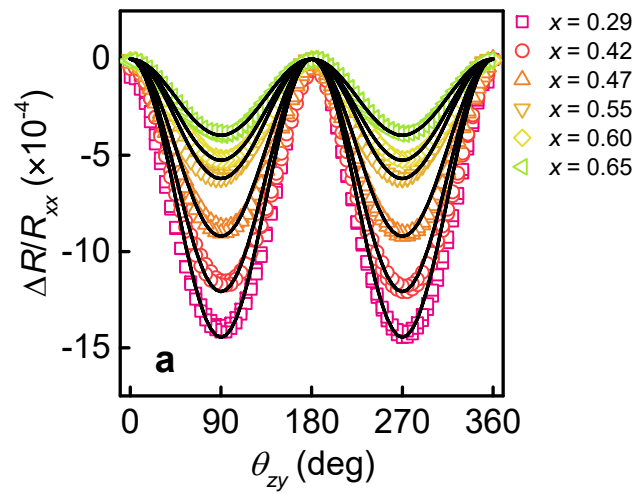
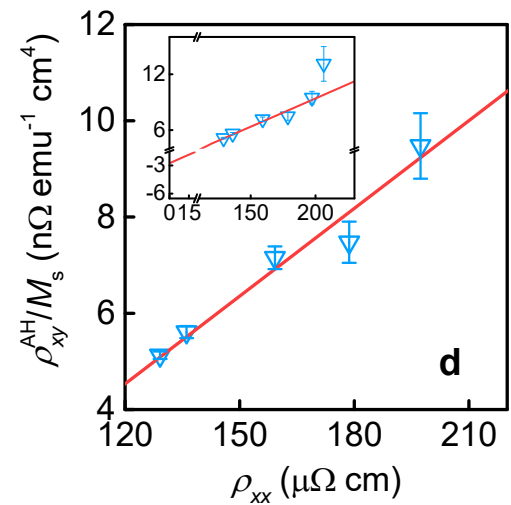
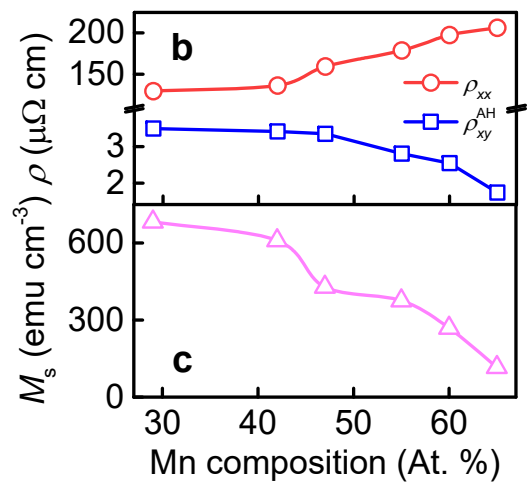
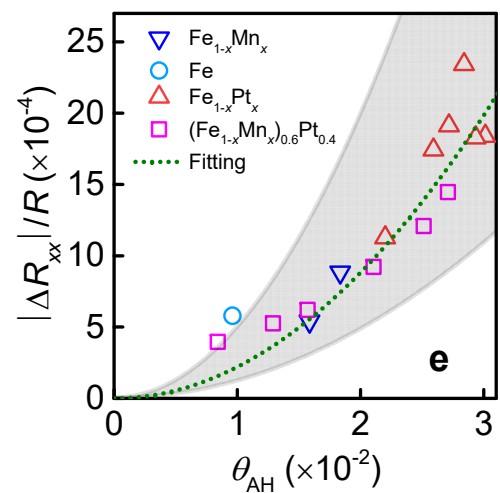

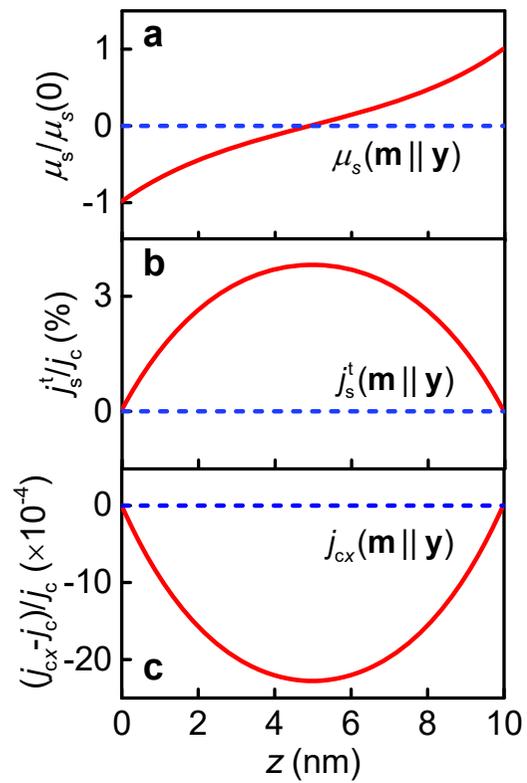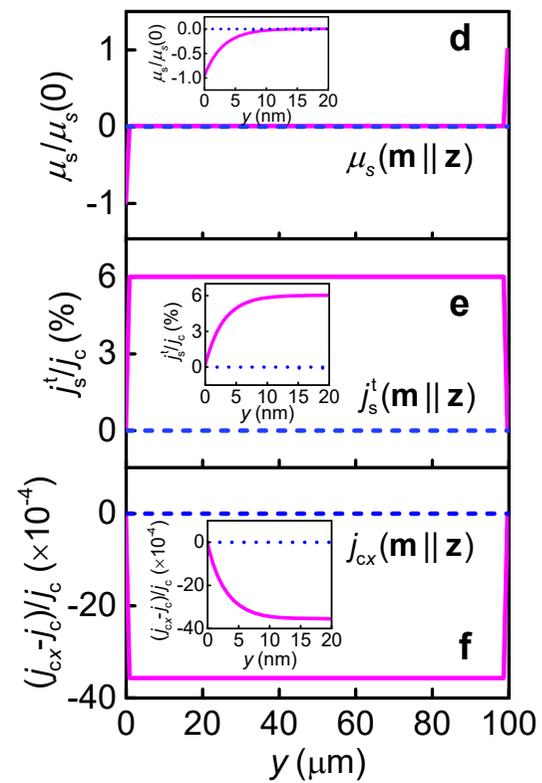

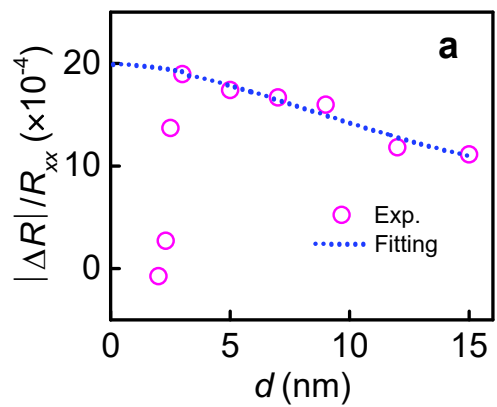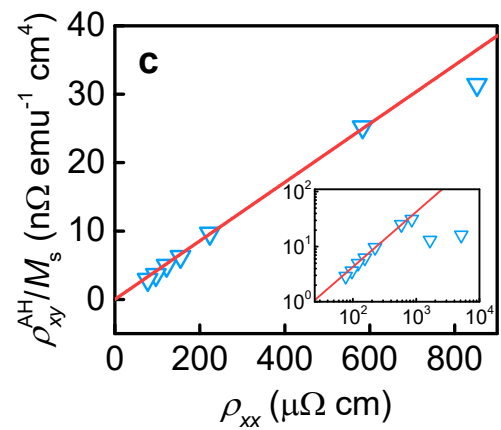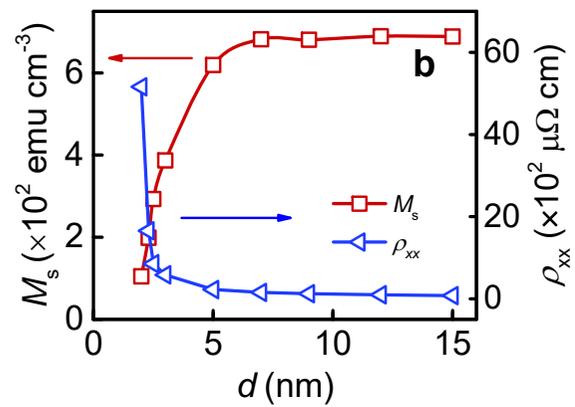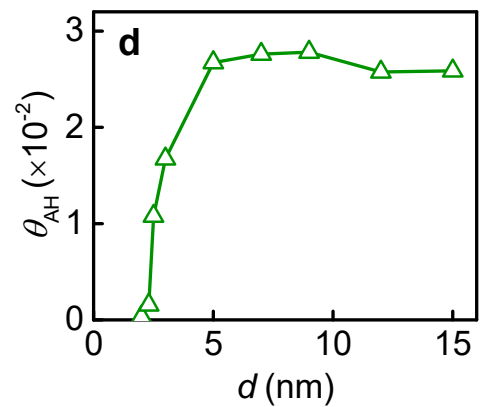

**Supplementary Information for "Anomalous Hall magnetoresistance in a ferromagnet"**

Yumeng Yang et al.

**Supplementary Note 1. Structural and magnetic properties of coupon films**

Supplementary Figure 1a shows the X-ray diffraction (XRD) patterns of four coupon films selected from the three types of FMs: [Fe$_{0.83}$Mn$_{0.17}$(0.6)/Pt(0.4)]$_{20}$ multilayer, (Fe$_{0.71}$Mn$_{0.29}$)$_{0.6}$Pt$_{0.4}$(20), Fe$_{0.71}$Mn$_{0.29}$(20) and Fe$_{0.4}$Mn$_{0.6}$(20), covering the peak range of fcc Pt (111) at 39.8°, fcc γ-Fe$_{1-x}$Mn$_x$ (111) at 43.5°, and bcc α-Fe$_{1-x}$Mn$_x$ (110) at 44.7°. To obtain moderate X-ray counts, the thickness of the films were kept at 20 nm, which is thicker than the patterned samples used for electrical measurements. It has been reported that Fe$_{1-x}$Mn$_x$ with $0.3 < x < 0.7$ is stable in γ-phase, while for $x < 0.3$ it undergoes a transition to α-phase with the increase of Fe composition[1,2]. The diffraction peak of Fe$_{0.71}$Mn$_{0.29}$ appears at 44.7°, suggesting that the film is dominantly in bcc α-phase (110), while the shift of the peak to 43.2° for Fe$_{0.4}$Mn$_{0.6}$ agrees with these previous reports that fcc γ-phase (111) becomes dominant when $x$ is below 0.3. On the other hand, the peaks at 40.9° - 41.1° observed in the multilayer and co-sputtered samples correspond to neither fcc Pt (111) at 39.8° nor Fe$_{1-x}$Mn$_x$ peaks. In the case of co-sputtered sample, it is understandable because Pt and Fe$_{1-x}$Mn$_x$ mixes uniformly to form an alloy. It is interesting to note that the diffraction peak for the multilayer appears at almost the same position as that of the co-sputtered film.

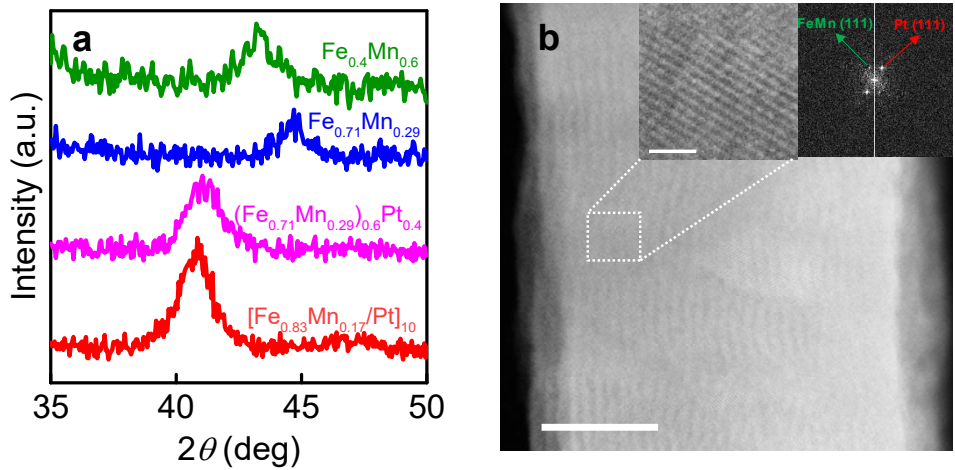

**Supplementary Figure 1. Structural characterizations of coupon films. a**, XRD pattern for [Fe$_{0.83}$Mn$_{0.17}$(0.6)/Pt(0.4)]$_{20}$, (Fe$_{0.71}$Mn$_{0.29}$)$_{0.6}$Pt$_{0.4}$(20), Fe$_{0.71}$Mn$_{0.29}$(20), and Fe$_{0.4}$Mn$_{0.6}$(20). **b,** STEM HAADF image of cross section of [Fe$_{0.5}$Mn$_{0.5}$(0.6)/Pt(0.6)]$_{30}$. Inset in **b** is an enlargement of the area enclosed by the dashed square and the FFT pattern of the same area. The white scale bar in **b** is 10 nm, and that in the inset is 1 nm.

To confirm the structure of the multilayer sample, we employed high resolution scanning transmission electron microscopy (STEM, a JEOL ARM200F) to directly image a multilayer sample consisting of [Fe$_{0.5}$Mn$_{0.5}$(0.6)/Pt(0.6)]$_{30}$. For a better recognition of the individual layer, the Pt thickness is slightly increased to be equal to that of FeMn, and the repetition period is also increased to 30. Supplementary Figure 1b is the cross-section high-angle annular dark field (HAADF) image of the sample, which shows a clear layer-by-layer structure except for some waviness of the layers. The HAADF image shows strong Z-contrast (Z being atomic number) and the Pt layers image bright. The waviness may come from the roughness of the SiO$_2$/Si substrate since no seed layer was deposited and the thickness of the individual layers is in the sub-nm range. As a result, it is difficult to separate and identify the individual layers clearly in the image. Fast Fourier transformation (FFT) was performed on the selected area enclosed by the dashed square. As shown in the inset of Supplementary Figure 1b, the FFT pattern has sharp spots, which correspond to fcc γ-Fe$_{0.5}$Mn$_{0.5}$ (111) and fcc Pt (111), respectively. This agrees with the above analysis based on XRD. The STEM and XRD results suggest that all the three types of samples are textured polycrystalline films, though the multilayer sample exhibits periodic structure with some waviness.

Supplementary Figures 2a – 2c show the in-plane $M$-$H$ loops for the three types of samples at room temperature: [Fe$_{1-x}$Mn$_x$(0.6)/Pt(0.4)]$_{10}$/Pt(1), (Fe$_{1-x}$Mn$_x$)$_{0.6}$Pt$_{0.4}$(9) and Fe$_{1-x}$Mn$_x$(9) with $x$ = 0.17 – 0.65. All the samples exhibit in-plane magnetic anisotropy as evident from the nearly square-like $M$-$H$ loops and small coercivity. The FM order gradually weakens with the increase of Mn composition, as reflected in the decrease of saturation magnetization $M_s$. As mentioned above, Fe$_{1-x}$Mn$_x$ undergoes a transition from bcc α-phase to fcc γ-phase with the increase of Mn composition. The former is known to be an FM phase whereas the latter is an antiferromagnet (AF)[1, 2]. Therefore, the decrease of $M_s$ can be understood as the gradual increase of AF ordering against the FM region as Mn composition increases. The inclusion of Pt in both the multilayer and co-sputtered samples extends the AF/FM boundary towards higher Mn

composition due to proximity effect between Pt and $Fe_{1-x}Mn_x$[3-5]. This is the reason why some of the co-sputtered and multilayer films still exhibit ferromagnetic properties in the range $x = 0.5 – 0.6$.

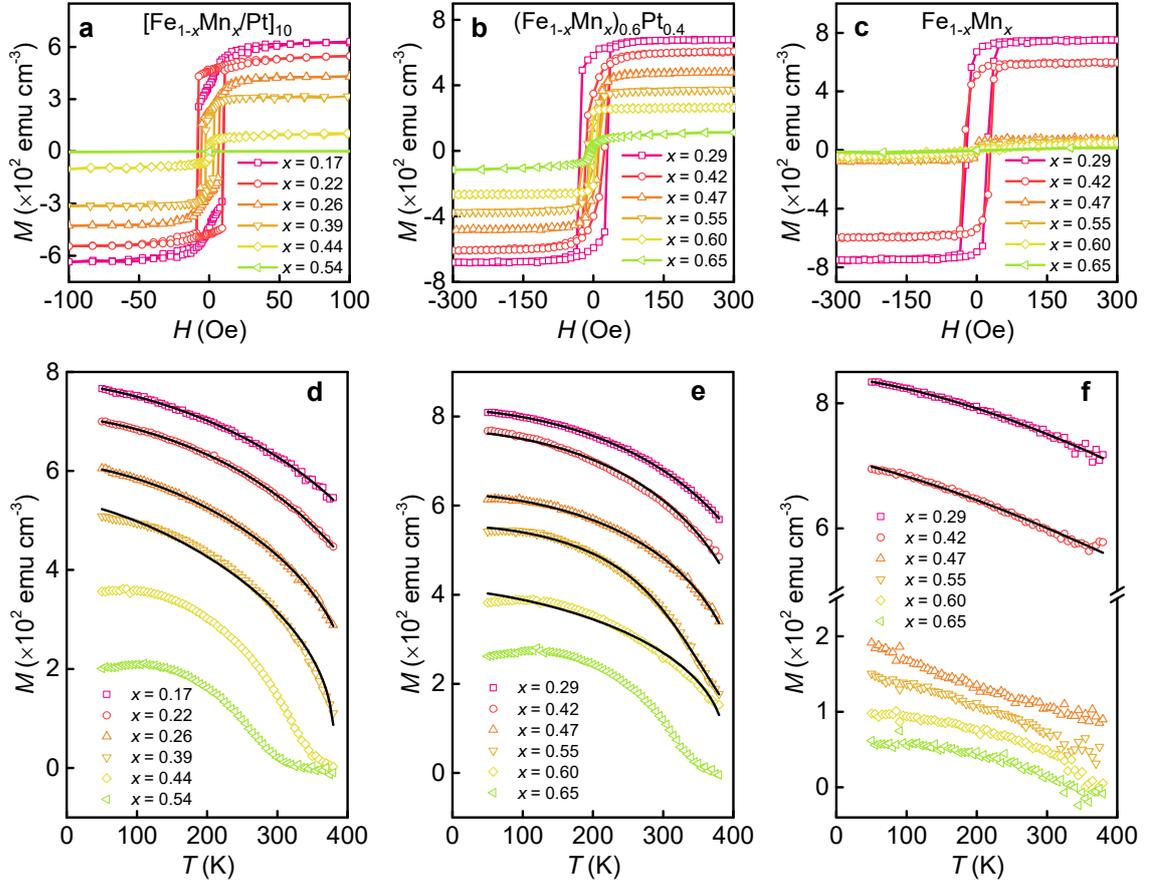

**Supplementary Figure 2. Magnetic properties of coupon films. a - c,** *M-H* loops: **a,** $[Fe_{1-x}Mn_x(0.6)/Pt(0.4)]_{10}/Pt(1)$; **b,** $(Fe_{1-x}Mn_x)_{0.6}Pt_{0.4}(9)$; **c,** $Fe_{1-x}Mn_x(9)$. **d - f,** *M-T* curves: **d,** $[Fe_{1-x}Mn_x(0.6)/Pt(0.4)]_{10}/Pt(1)$; **e,** $(Fe_{1-x}Mn_x)_{0.6}Pt_{0.4}(9)$; **f,** $Fe_{1-x}Mn_x(9)$. Solid lines in **d - f** are fittings based on Supplementary Equation 1.

To gain more insight into the magnetic properties, we examined the temperature dependence of magnetization in these samples, and the results are summarized in Supplementary Figures 2d – 2f. As with most ferromagnetic materials, the $M_s$ decreases with the increase of temperature. It is also apparent that the Curie temperature ($T_C$) decrease with increasing Mn composition. For a more quantitative understanding, we invoke the semi-empirical model developed by Kuz'min[6, 7], which turned out to be very successful in fitting the *M-T* curves of many different types of magnetic materials, to fit these curves in Supplementary Figures 2d – 2f. According to this model, the temperature dependent magnetization of FM is given by

$$M(T) = M(0)\left[1 - s\left(\frac{T}{T_C}\right)^{3/2} - (1-s)\left(\frac{T}{T_C}\right)^{5/2}\right]^b \quad (1)$$

where $M(0)$ is the magnetization at $T = 0$ K, $T_C$ is the Curie temperature, $s$ is the so-called shape parameter with a value in the range of 0 - 2.5, and $b$ is the critical exponent whose value is determined by the universality class of the material: 0.125 for two-dimensional Ising, 0.325 for three-dimensional (3D) Ising, 0.346 for 3D XY, 0.365 for 3D Heisenberg, and 0.5 for mean-field theory[8]. On the other hand, for surface magnetism, $b$ is in the range of 0.75–0.89[9, 10]. Considering the 3D ferromagnetic nature of these samples, we fixed $b$ at 0.365 and fitted the $M$-$T$ curves by optimizing the remaining parameters. The values are summarized in Supplementary Table 1. The fitting is generally good for samples with low Mn compositions (around $x = 0.4$ for FeMn and FeMn/Pt multilayers and $x = 0.6$ for FeMnPt alloy). However, at higher Mn compositions, the $M$-$T$ curves tend to deviate from that described by Supplementary Equation 1 with the $s$ value out of its normal range; therefore we did not fit these curves and leave the parameters as N.A. in the table. The deviation is presumably caused by the onset of AFM ordering, and therefore, understandably, their $M$-$T$ curves will not follow that of the FM. According to M. D. Kuz`min *et al.*, for 3D Heisenberg magnets, $s$ is determined by the dependence of exchange interaction on interatomic distance[7]. It is generally positive with a small $s$ (< 0.4) corresponding to metallic FMs with long-range ferromagnetic ordering and high $T_C$, whereas a large $s$ (> 0.8) is indicative of competing exchange interactions and the resultant material typically has a low $T_C$. Following this reasoning, among the three types of samples, the co-sputtered samples behave more like a metallic FM than the other two types of samples do, and therefore, in the main text, the analysis on AHE and MR data have been focused on the co-sputtered samples.

Supplementary Table 1. Summary of the fitting parameters using Supplementary Equation 1 for the three types of samples. N.A. is the abbreviation for not available.

| Type | $x$ | $M(0)$ (emu cm$^{-3}$) | $s$ | $T_C$ (K) |
|---|---|---|---|---|
| Fe$_{1-x}$Mn$_x$/Pt | 0.17 | 774 | 0.96 | 516 |
| | 0.22 | 709 | 0.93 | 468 |
| | 0.26 | 613 | 1.07 | 417 |
| | 0.39 | 536 | 1.48 | 382 |
| | 0.44 | N.A. | N.A. | N.A. |
| | 0.54 | N.A. | N.A. | N.A. |
| (Fe$_{1-x}$Mn$_x$)$_{0.6}$Pt$_{0.4}$ | 0.29 | 815 | 0.47 | 481 |
| | 0.42 | 767 | 0.45 | 441 |
| | 0.47 | 626 | 0.45 | 419 |
| | 0.55 | 576 | 1.10 | 386 |
| | 0.60 | 410 | 1.07 | 392 |
| | 0.65 | N.A. | N.A. | N.A. |
| Fe$_{1-x}$Mn$_x$ | 0.29 | 840 | 1.38 | 844 |
| | 0.42 | 703 | 1.20 | 650 |
| | 0.47 | N.A. | N.A. | N.A. |
| | 0.55 | N.A. | N.A. | N.A. |
| | 0.60 | N.A. | N.A. | N.A. |
| | 0.65 | N.A. | N.A. | N.A. |

**Supplementary Note 2. Field dependent magnetoresistance (FDMR) measurement results**

As a supplementary reference, FDMR results are shown in Supplementary Figures 3a – 3l for the [Fe$_{0.83}$Mn$_{0.17}$(0.6)Pt(0.4)]$_{10}$/Pt(1) multilayer, co-sputtered (Fe$_{0.71}$Mn$_{0.29}$)$_{0.6}$Pt$_{0.4}$(9), Fe$_{0.71}$Mn$_{0.29}$(9) and Fe(9) with applied sweeping field in $x$-, $y$- and $z$-axis, respectively. The data are thus denoted as $H_x$, $H_y$, and $H_z$ FDMR curves, respectively. As can be seen from Supplementary Figures 3a, 3d, 3g and 3j, the shape of the $H_x$ and $H_y$ FDMR curves resembles that of the conventional AMR, which gives $\rho_x > \rho_y$ with $\rho_x$ ($\rho_y$) the resistivity when $\mathbf{m} \parallel \mathbf{x}$ ($\mathbf{m} \parallel \mathbf{y}$). For the $H_z$ FDMR curves, the negative MR above saturation field (around 15 kOe) is attributed to the so called spin disorder MR[11]. Below saturation field, depending on the $H_z$ field misalignment direction, two different shapes can be obtained. When $H_z$ is misaligned from z-axis towards x-axis, the additional x-component of field helps rotate the magnetization in zx plane, and an M shaped MR curve is observed (Supplementary Figures 3b, 3e, 3h, 3k). On the other hand, when it is misaligned from z-axis towards y-axis, the magnetization is rotated in zy plane by the additional y-component, and a W shaped MR curve is observed (Supplementary Figures 3c, 3f, 3i, 3l). In view of these shapes and the magnetization positions, it can be inferred that $\rho_x > \rho_z > \rho_y$ with $\rho_z$ the resistivity when $\mathbf{m} \parallel \mathbf{z}$. This relation is in agreement with the above $H_x$ and $H_y$ FDMR curves and ADMR results in Fig. 2 of the main text. To support the explanation, we performed macro-spin simulation for the $H_z$ case following the approach described in our previous work[3]. Taking into consideration the misalignment, the applied field **H** is expressed as:

$$\mathbf{H} = (H \sin\delta \cos\chi, H \sin\delta \sin\chi, H \cos\delta) \qquad (2)$$

here $\delta$ and $\chi$ are the misalignment polar and azimuth angles, respectively. The free energy density (normalized to saturation magnetization) is given by

$$E = \frac{H_d}{2}\cos^2\theta - H(\sin\theta\cos\varphi\sin\delta\cos\chi + \sin\theta\sin\varphi\sin\delta\sin\chi + \cos\theta\cos\delta) - \frac{H_k}{2}\sin^2\theta\cos^2\varphi \qquad (3)$$

where $\theta$ and $\varphi$ are the polar and azimuth angles of the magnetization, respectively; $H_k = 2K_u/M_s$ is the anisotropy field; and $H_d$ is the demagnetizing field. Here, due to the sample geometry, the easy axis is assumed to be in x-axis, parallel to the current direction. Supplementary Equation 3 can be solved numerically to obtain the equilibrium angle $\varphi$ and $\theta$ as a function of $H$.

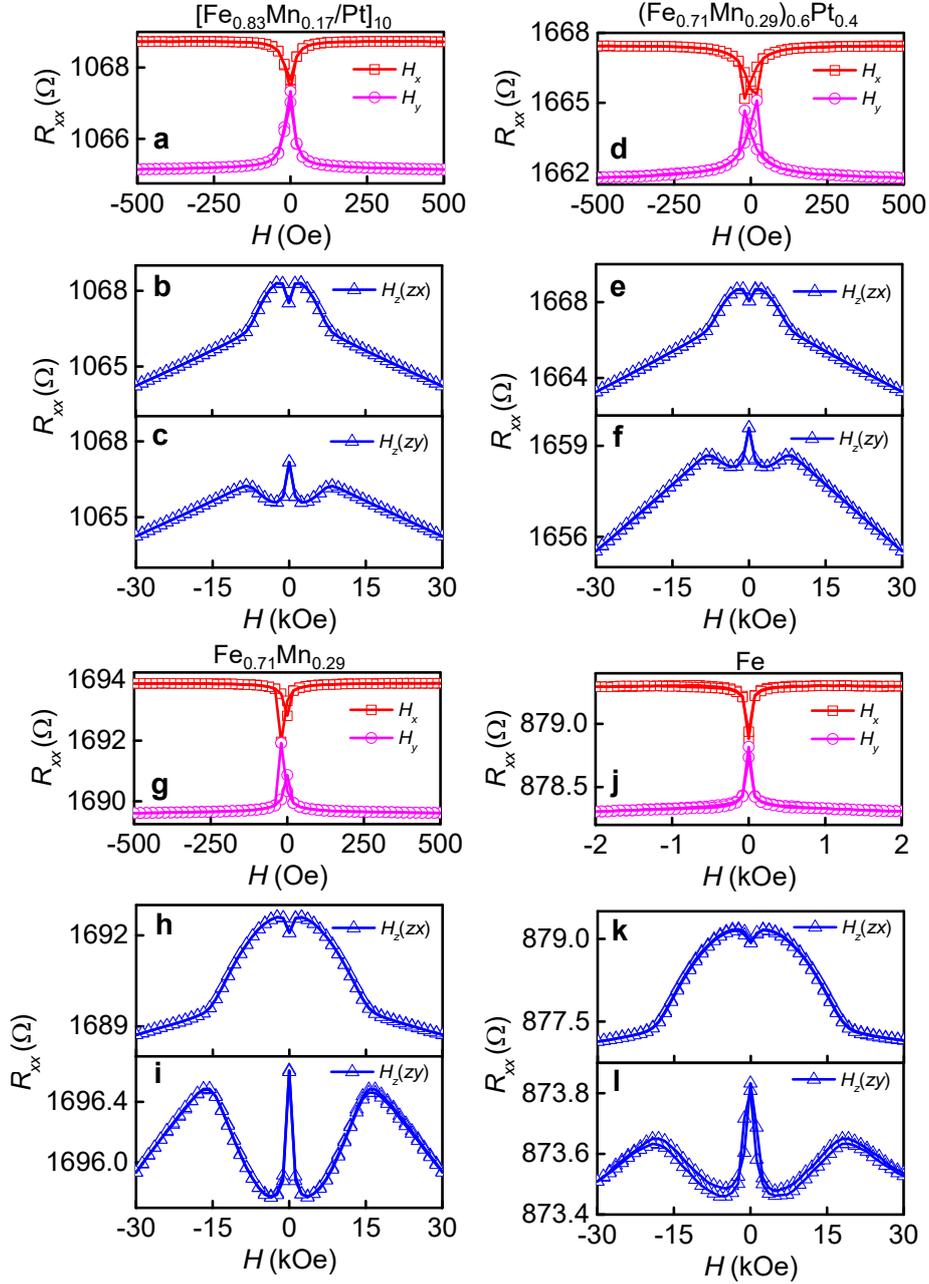

**Supplementary Figure 3. FDMR curves for different types of samples. a - c,** [Fe$_{0.83}$Mn$_{0.17}$(0.6)/Pt(0.4)]$_{10}$/Pt(1); **d - f,** (Fe$_{0.71}$Mn$_{0.29}$)$_{0.6}$Pt$_{0.4}$(9); **g - i,** Fe$_{0.71}$Mn$_{0.29}$(9); **j - l,** Fe(9). The legends $H_x$, $H_y$ and $H_z$ denotes the FDMR curves obtained when the field is swept in $x$, $y$, and $z$-axis direction, respectively; and $zx$ (or $zy$) in the parenthesis after $H_z$ indicates the misalignment of $H_z$ from $z$-axis towards $x$-axis (or $y$-axis).

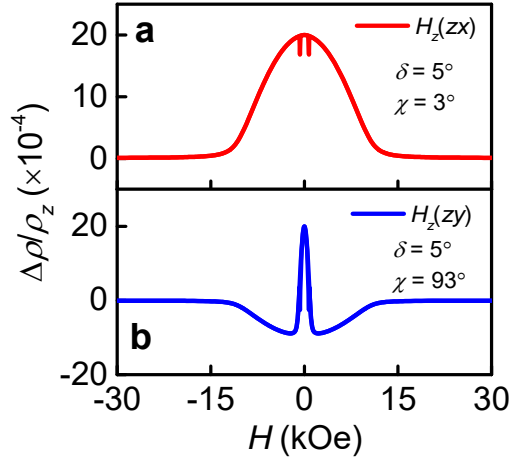

**Supplementary Figure 4. Simulated $H_z$ FDMR curves with $\rho_x > \rho_z > \rho_y$ and the small field misalignment. a,** $H_z$ is misaligned from z-axis towards x-axis with $\delta = 5°$, $\chi = 3°$; **b,** $H_z$ is misaligned from z-axis towards y-axis with $\delta = 5°$, $\chi = 93°$.

Taking into account the relation $\rho_x > \rho_z > \rho_y$, the MR ratio can be written as:

$$\frac{\Delta\rho}{\rho_z} = \frac{\rho_x - \rho_z}{\rho_z}\sin^2\theta\cos^2\varphi + \frac{\rho_y - \rho_z}{\rho_z}\sin^2\theta\sin^2\varphi \tag{4}$$

By using $\frac{\rho_x - \rho_z}{\rho_z} = 2\times 10^{-3}$ and $\frac{\rho_y - \rho_z}{\rho_z} = -1\times 10^{-3}$ as estimated from ADMR results in the main text, the $H_z$ FDMR curves can be reproduced in Supplementary Figures 4a and 4b. The parameters used are: $H_d = 10$ kOe, $H_k = 50$ Oe and $\delta = 5°$, $\chi = 3°$ for Supplementary Figure 4a (or $\delta = 5°$, $\chi = 93°$ for Supplementary Figure 4b). As can been from the figure, the M and W shaped MR curves can be reproduced well as long as a small misalignment of $H$ from the z-axis and $\rho_x > \rho_z > \rho_y$ are assumed. The results from control samples, exhibiting different features, will be discussed in Supplementary Note 6 shortly.

**Supplementary Note 3. Demagnetizing field effect on angle dependent magnetoresistance (ADMR) results**

As discussed in the main text, the fittings of ADMR results in Figs. 2b – 2e exhibit a small deviation from the $\sin^2\theta_{zx}$ or $-\sin^2\theta_{zy}$ dependence, especially in the case of Fe. This is caused by the small deviation of the magnetization with respect to the external field during *zx*- or *zy*-plane rotation in the presence of a moderate demagnetizing field ($H_d$). In a similar way, Supplementary Equation 3 can be used to numerically calculate the equilibrium angle $\varphi$ and $\theta$ during the field rotation. Here, we look at the *zy*-plane rotation case as an example, and the *zx*-plane rotation case is similar. We let $\chi = 90°$, $H = 30$ kOe and vary $\delta$ from 0 to 360°. From the AHE measurements in Supplementary Figure 5a, $H_d$ in [Fe$_{0.83}$Mn$_{0.17}$(0.6)Pt(0.4)]$_{10}$/Pt(1), (Fe$_{0.71}$Mn$_{0.29}$)$_{0.6}$Pt$_{0.4}$(9), Fe$_{0.71}$Mn$_{0.29}$(9) and Fe(9) can be estimated as 10 kOe, 10 kOe, 15 kOe and 20 kOe, respectively. Supplementary Figure 5b shows the calculated $\theta$ angle during the *zy*-plane rotation, corresponding to different $H_d$ values. As can be seen, although the magnetization can be aligned in the *z*-axis ($H_d < 30$ kOe), the magnetization indeed presents a small deviation from the perfect alignment at other angles, and this deviation increases as $H_d$ increases. This directly causes the deviation of ADMR fittings as manifested in the $-\sin^2\theta$ calculation in Supplementary Figure 5c. It is clear that these numerical results are in good agreement with the measured ADMR results in Fig. 2 of the main text. However, it should be noted that the presence of the magnetization deviation from the external field does not affect the magnitude of the MR ratio. As long as the external field is above the saturation field, the magnetization is able to reach *z*- and *y*-axis in equilibrium, and therefore, it does not affect the discussion based on the MR ratios.

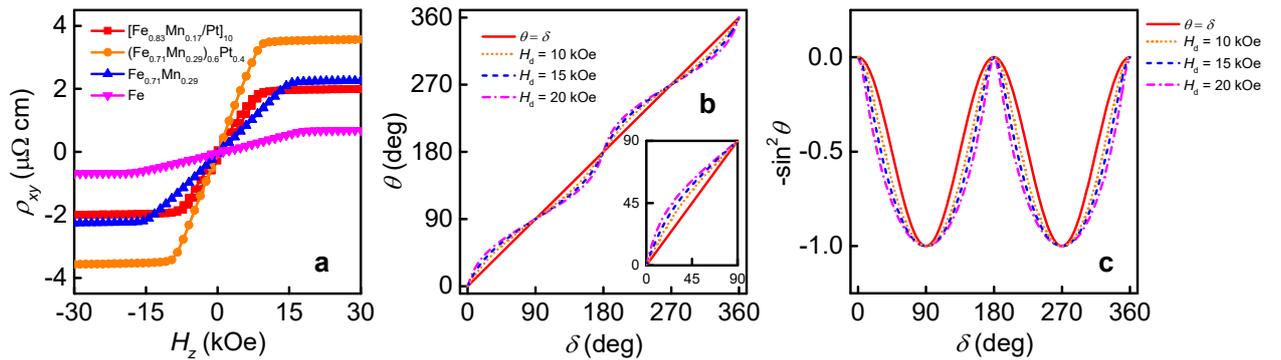

**Supplementary Figure 5. Simulated ADMR curve in *zy*-plane rotation. a,** Hall resistivity as a function of $H_z$ in [Fe$_{0.83}$Mn$_{0.17}$(0.6)Pt(0.4)]$_{10}$/Pt(1), (Fe$_{0.71}$Mn$_{0.29}$)$_{0.6}$Pt$_{0.4}$(9), Fe$_{0.71}$Mn$_{0.29}$(9) and Fe(9); **b,** Simulated magnetization angle $\theta$ during *zy*-plane rotation with different $H_d$ values; **c,** Simulated $-\sin^2\theta$ dependence using the $\theta$ data in **b.**

**Supplementary Note 4. Correlation of MR($\theta_{zy}$) with anomalous Hall effect (AHE) in Fe$_{1-x}$Mn$_x$/Pt multilayers and Fe$_{1-x}$Mn$_x$**

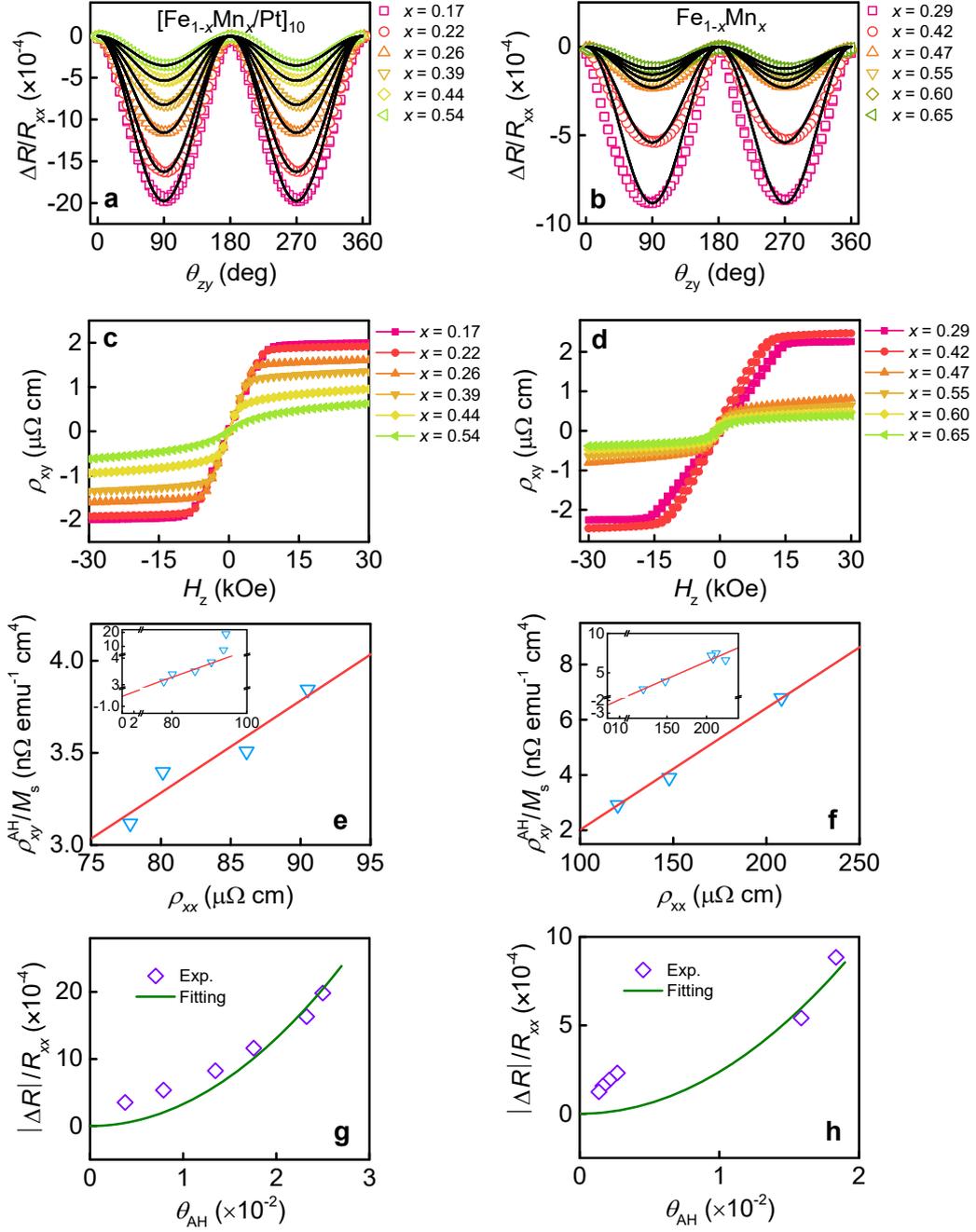

**Supplementary Figure 6. Correlation of MR($\theta_{zy}$) and AHE. a** and **b,** MR($\theta_{zy}$) ratio for [Fe$_{1-x}$Mn$_x$(0.6)/Pt(0.4)]$_{10}$/Pt(1) and Fe$_{1-x}$Mn$_x$(9), respectively, with $x = 0.17 - 0.65$; **c** and **d,** Hall resistivity as a function of $H_z$. **e** and **f,** $\rho_{xy}^{AH} / M_s$ as a function of $\rho_{xx}$ in the linear range. **g** and **h,** Plot of MR ratio as a function of $\theta_{AH}$ and fitting using the quadratic relation $\theta_{AH}^2$. Insets in **e** and **f** are the full range of the plot in the respective figure, and solid lines in **e** and **f** serve as a guide for the eye.

Supplementary Figures 6a and 6b show the MR($\theta_{zy}$) curves for [Fe$_{1-x}$Mn$_x$(0.6)/Pt(0.4)]$_{10}$/Pt(1) and Fe$_{1-x}$Mn$_x$(9), respectively, with $x = 0.17 – 0.65$. Similar to the results shown in Fig. 3 of the main text for co-sputtered (Fe$_{1-x}$Mn$_x$)$_{0.6}$Pt$_{0.4}$(9) samples, the MR ratio in these two types of samples also decreases with the increase of Mn composition. In addition, the decreasing trend happens to coincide with the experimentally determined $M_s$ dependence on Mn composition (see Supplementary Figures 2a and 2c). Supplementary Figures 6c and 6d show the Hall measurement results for Fe$_{1-x}$Mn$_x$/Pt multilayer and Fe$_{1-x}$Mn$_x$ samples with field applied perpendicular to the plane. Similar to the procedures described in the main text for the co-sputtered samples, we extracted $\rho_{xy}^{AH}$ from the measured Hall resistivity and then plot $\rho_{xy}^{AH} / M_s$ as a function of $\rho_{xx}$ in Supplementary Figures 6e and 6f for [Fe$_{1-x}$Mn$_x$(0.6)/Pt(0.4)]$_{10}$/Pt(1) and Fe$_{1-x}$Mn$_x$(9), respectively. Interestingly, the linear fitting still holds in both cases for samples in the low Mn composition range, whose *M-T* curves can be fitted by Supplementary Equation 1. At high Mn compositions, due to the weakening of FM order and onset of AF order, the fitting significantly deviates from the linear relation, particularly for the multilayer sample. As with the case of co-sputtered samples, $\theta_{AH}$ can be calculated from $\theta_{AH} = \rho_{xy}^{AH} / \rho_{xx}$. With these values, we can plot MR($\theta_{zy}$) ratio as a function of $\theta_{AH}$, as shown in Supplementary Figures 6g and 6h, respectively. Interestingly, for Fe$_{1-x}$Mn$_x$/Pt multilayers, the quadratic relation between MR($\theta_{zy}$) and $\theta_{AH}$ still holds approximately including samples with high Mn composition. This is not surprising because the saturation magnetization does not appear explicitly in the drift-diffusion formalism, though it may indirectly affect the scattering asymmetry parameters. Further studies are required to gain an insight of the role of $M_s$ in determining the MR($\theta_{zy}$) versus $\theta_{AH}$ relation. On the other hand, the fitting for Fe$_{1-x}$Mn$_x$ only serves as guide for eye due to the small number of data points. As mentioned above, the magnetic properties of Fe$_{1-x}$Mn$_x$ changes drastically when $x$ approaches and exceeds 0.5. A more rigorous theoretical model is required to deal with AHE of such kind of materials with complex spin structures. Before ending this supplementary note, it is worth pointing out that, in the case of multilayers, MR($\theta_{zy}$) may also originate from SHE in the individual Pt layers[12] or interface scattering[13], as we reported previously[3, 4]. However, it is difficult to distinguish AHE and SHE contributions to MR($\theta_{zy}$) as both exhibit the same angular dependence.

**Supplementary Note 5. Determination of anomalous Hall resistivity from Hall measurements**

Hall resistivity in FM metals can be empirically written as[14]: $\rho_{xy} = R_0 H_z + R_s M(H_z)$, where $R_0$ and $R_s$ are the ordinary Hall effect (OHE) and AHE coefficients, respectively, $H_z$ is the applied magnetic field in $z$-direction, and $M(H_z)$ is the magnetization at $H_z$. The OHE arises from the Lorentz force generated by the magnetic field, which goes to zero when $H_z = 0$. Therefore, in order to extract the AHE contribution $\rho_{xy}^{AH}$, $\rho_{xy}$ is extrapolated from positive and negative high fields to zero field as indicated by the solid line in Supplementary Figure 7. Here, the Hall result for (Fe$_{1-x}$Mn$_x$)$_{0.6}$Pt$_{0.4}$(9) is used as an example. The difference of the two intercepts (positive and negative) at $H_z = 0$ corresponds to $2\rho_{xy}^{AH}$. On the other hand, the slope of the solid line corresponds to $R_0$. By repeating the above process, $\rho_{xy}^{AH}$ in different samples were obtained.

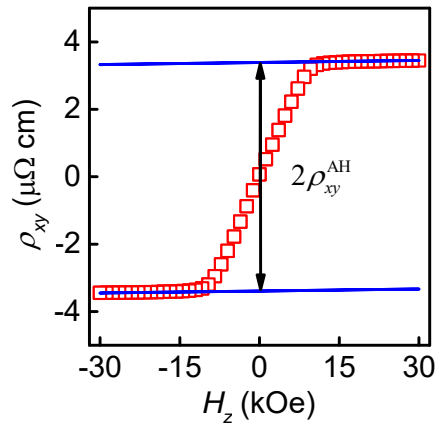

**Supplementary Figure 7. Extraction of $\rho_{xy}^{AH}$ from Hall measurements.** An example illustrates the process of extracting $\rho_{xy}^{AH}$ from Hall measurement results.

**Supplementary Note 6. Control measurements on Co, Py and $Ir_{0.2}Mn_{0.8}$ thin films**

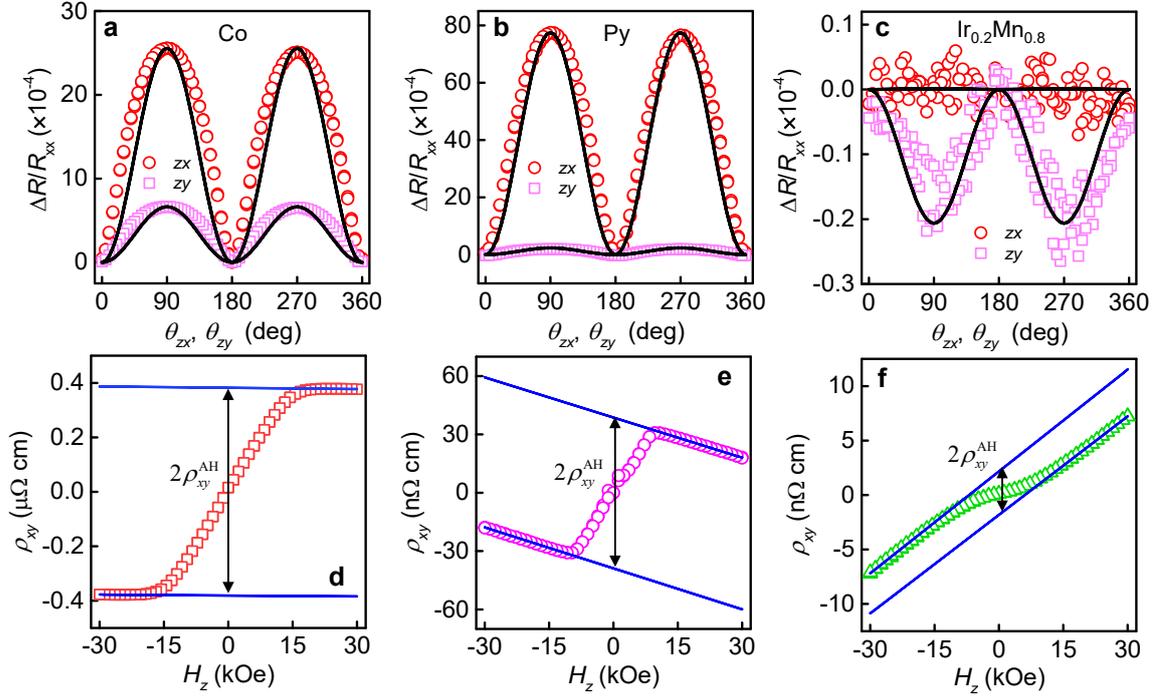

**Supplementary Figure 8. Investigation of AHMR in Co, Py and $Ir_{0.2}Mn_{0.8}$ thin films. a – c,** ADMR measurement results: **a,** Co; **b,** Py; **c,** $Ir_{0.2}Mn_{0.8}$. **d – f,** Hall measurement results: **d,** Co; **e,** Py; **f,** $Ir_{0.2}Mn_{0.8}$. Solid lines in **a - c** are the fittings based on the angle dependence; and those in d – f are the linear fitting to extract $\rho_{xy}^{AH}$ using the method described in Supplementary Note 5.

In addition to Fe based materials, we performed the same magnetoresistance measurements for control samples including Co(9), Py(9) and $Ir_{0.2}Mn_{0.8}$(9) thin films. Supplementary Figures 8a – 8c show the MR results for *zx* and *zy* plane scans in these samples, respectively. And shown in Supplementary Figures 8d – 8f are the respective Hall measurement results. Following the discussion in the main text, MR($\theta_{zx}$) corresponds to the AMR while MR($\theta_{zy}$) is from the AHMR. As expected, a sizable AMR [MR($\theta_{zx}$) signal] can be observed in both Co and Py samples, and it is vanished in $Ir_{0.2}Mn_{0.8}$. On the other hand, the MR($\theta_{zy}$) signal in Co and Py exhibits an opposite polarity with that expected for AHMR, which gives $\rho_y > \rho_z$. Previous reports suggested that such kind of behavior may come from the geometric size effect (GSE)[15-17], which itself is still debatable as different mechanisms have been suggested such as electronic structure in thin films[15], anisotropic interfacial scattering[16] and anisotropic *sd* scattering of minority spins[17]. Similar observation of $\rho_x > \rho_y > \rho_z$ has been observed in Ni[15], Py[16] and Co[17] in these previous reports. Put the origin aside first, our results suggested that AHMR in Co and Py samples are much smaller than those from Fe-based samples and are masked out by GSE related MR. For $Ir_{0.2}Mn_{0.8}$, although the polarity of MR($\theta_{zy}$) agrees with the prediction of AHMR, its size is around two orders of magnitude smaller than that observed in $Fe_{1-x}Mn_x$ based systems. This is understandable because although a relatively large spin Hall angle, around 80% of that of Pt, has been reported for $Ir_{0.2}Mn_{0.8}$[18, 19], as an AF, it does not have a net moment which is required for observing AHE.

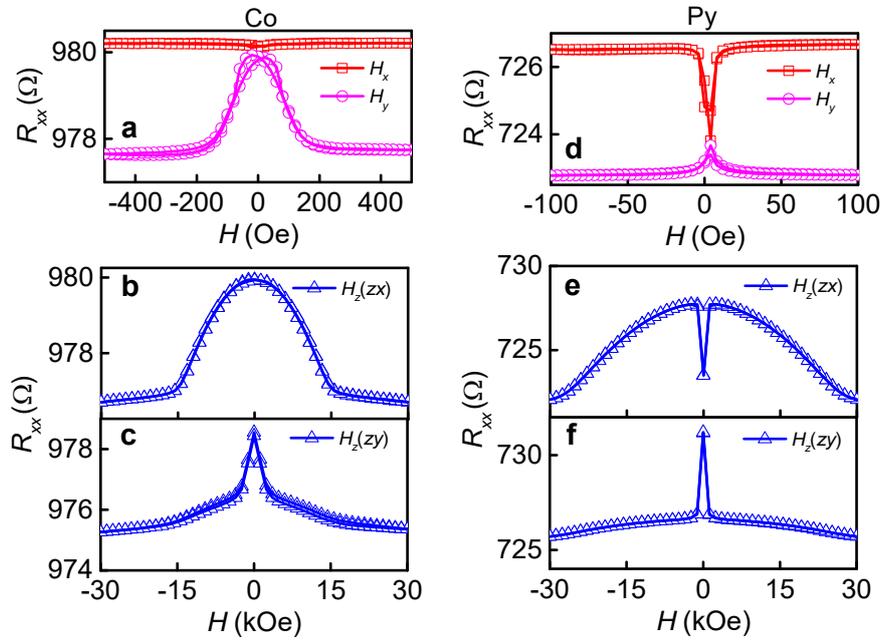

**Supplementary Figure 9. FDMR curves for control samples. a – c,** Co(9); **d – f,** Py(9). The legends $H_x$, $H_y$ and $H_z$ denotes the FDMR curves obtained when the field is swept in *x*, *y*, and *z*-axis direction, respectively; and *zx* (or *zy*) in the parenthesis after $H_z$ indicates the misalignment of $H_z$ from *z*-axis towards *x*-axis (or *y*-axis).

The FDMR measurements were also performed for Co and Py to confirm the different relation of $\rho_x > \rho_y > \rho_z$ as compared to the Fe based samples. The results are shown in Supplementary Figures 9a – 9c for Co and Supplementary Figures 9d – 9f for Py with different sweeping field directions. The general shape of $H_x$ and $H_y$ FDMR curves (Supplementary Figures 9a and 9d) are similar to the previous ones in Supplementary Note 2, which gives $\rho_x > \rho_y$. The different magnitude of the dip or peak in the center small field region is related to the detailed domain structure formed during the sweeping[17]. The $H_z$ FDMR curves with misalignment to *x*-axis also looks similar to those in Supplementary Note 2, but those with misalignment to *y*-axis exhibit a totally different shape. This is directly related to $\rho_y > \rho_z$. The same macro-spin approach can be applied to the Co and Py case as well to determine $\theta$ and $\varphi$. The difference is that in the MR ratio simulation, a positive ratio of $\frac{\rho_y - \rho_z}{\rho_z} = 1 \times 10^{-3}$ should be adopted. For illustration purpose, we use the same set of parameters to simulate the MR curves, and the results are presented in Supplementary Figures. 10a and 10b. As can be seen, indeed the shapes can be accounted for by the assumption of a small misalignment of *H* from the *z*-axis and $\rho_x > \rho_y > \rho_z$. Both the FDMR and ADMR results suggest that the MR effects, especially MR($\theta_{zy}$) in Fe based samples are different from those in Co and Py.

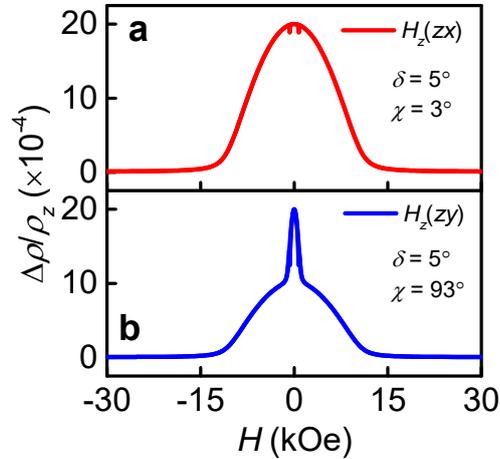

**Supplementary Figure 10. Simulated $H_z$ FDMR curve with $\rho_x > \rho_y > \rho_z$ and the small field misalignment. a,** $H_z$ is misaligned from *z*-axis towards *x*-axis with $\delta = 5°$, $\chi = 3°$; **b,** $H_z$ is misaligned from *z*-axis towards *y*-axis with $\delta = 5°$, $\chi = 93°$.

The difference in MR prompted us to look into the AHE in these samples. In general, the Hall resistivity in these control is at least one order of magnitude smaller than that in Fe based systems with the same thickness. By using the method described in Supplementary Note 5, we separated the contribution of OHE and AHE from the Hall effect results. The values are summarized in Supplementary Table 2 and compared with those obtained from Fe based systems. It should be noted that based on early theoretical calculations on intrinsic AHE[20, 21], both OHE and AHE resistivity are negative for Ni, while both of them are positive for Fe; and for Co, OHE is negative but AHE is positive. Here, the positive resistivity refers to the positive Hall voltage in the positive *y*-axis direction when the current is in the positive *x*-axis direction and magnetic field is in positive *z*-axis direction. As can be seen from the table, the signs of the OHE and AHE contributions follow the calculations, if one considers the fact that Fe is the major composition in these samples, except for $\rho_{xy}^{AH}$ in Py. In fact, it has been pointed out that due to the small strength of AHE, the sign of $\rho_{xy}^{AH}$ in Py is very sensitive to chemical composition as well as the thickness of film, and therefore both positive and negative values has been reported in Py[22]. From these values, it is clear that smaller $\theta_{AH}$ is the direct cause of the difference in MR($\theta_{zy}$) between Co, Py and Ir$_{0.2}$Mn$_{0.8}$ and Fe-based samples.

Supplementary Table 2. Comparison of the OHE, AHE resistivity and AHE angle among Fe$_{0.83}$Mn$_{0.17}$/Pt, (Fe$_{0.71}$Mn$_{0.29}$)$_{0.6}$Pt$_{0.4}$, Fe$_{0.71}$Mn$_{29}$, Fe$_{0.75}$Pt$_{25}$, Fe, Co, Py and Ir$_{0.2}$Mn$_{0.8}$. The thickness of these samples are fixed at 9 nm.

| Type | $\rho_0$ (μΩ cm) | $\rho_{xy}^{AH}$ (μΩ cm) | $\theta_{AH}$ |
|---|---|---|---|
| Fe$_{0.83}$Mn$_{0.17}$/Pt | 0.12±0.02 | 1.94±0.01 | 0.025±7.55×10$^{-5}$ |
| (Fe$_{0.71}$Mn$_{0.29}$)$_{0.6}$Pt$_{0.4}$ | 0.15±0.03 | 3.38±0.04 | 0.028±1.60×10$^{-4}$ |
| Fe$_{0.71}$Mn$_{29}$ | 0.12±0.01 | 2.21±0.05 | 0.018±1.75×10$^{-4}$ |
| Fe$_{0.75}$Pt$_{25}$ | 0.12±0.02 | 1.61±0.02 | 0.022±1.14×10$^{-4}$ |
| Fe | 0.07±0.01 | 0.65±0.01 | 0.0096±5.69×10$^{-5}$ |
| Co | -0.009±0.002 | 0.38±0.001 | 0.0037±9.32×10$^{-6}$ |
| Py | -0.05±0.001 | 0.04±0.0003 | 5.71×10$^{-4}$±4.11×10$^{-6}$ |
| Ir$_{0.2}$Mn$_{0.8}$ | 0.02±0.001 | -0.002±0.0002 | -7.30×10$^{-6}$±8.27×10$^{-7}$ |

**Supplementary Note 7. Discussion on magnetic field misalignment and geometric size effect (GSE) related anisotropic magnetoresistance (AMR) contributions**

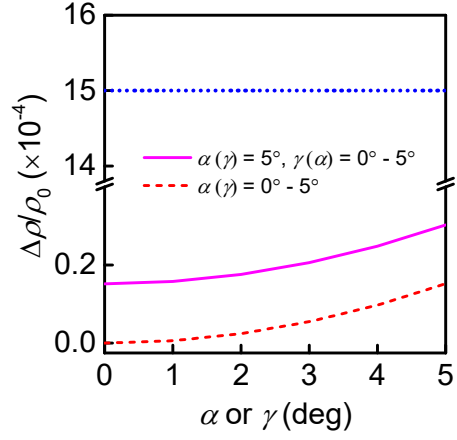

**Supplementary Figure 11. Simulated AMR contribution to MR($\theta_{zy}$) ratio due to field misalignment.** Solid line: $\alpha = 5°$ and $\gamma = 0°$ to $5°$ or $\alpha = 0°$ to $5°$ and $\gamma = 5°$. $\alpha = 0°$ and $\gamma = 0°$ to $5°$ or $\alpha = 0°$ to $5°$ and $\gamma = 0°$. Dotted line: experimental MR($\theta_{zy}$) ratio for $(Fe_{0.71}Mn_{0.29})_{0.6}Pt_{0.4}$ with a thickness of 9 nm.

Before proceeding further, it is necessary to exclude other contributions besides AHE as the main cause for the MR($\theta_{zy}$). In this note, we first discuss the influence of field misalignment contribution. During the $zy$ plane ADMR measurement, the misalignment of either field or sample (which are relative to each other) can be represented by a small rotation around $y$-axis by $\gamma$ and $z$-axis by $\alpha$. Assume that, at perfect alignment, the magnetization vector is given by:

$$\mathbf{m} = \begin{pmatrix} 0 \\ \sin\theta_{zy} \\ \cos\theta_{zy} \end{pmatrix} \tag{5}$$

After the rotation around $y$- and $z$-axis, the magnetization vector is given by:

$$\mathbf{m} = \begin{pmatrix} \cos\alpha & -\sin\alpha & 0 \\ \sin\alpha & \cos\alpha & 0 \\ 0 & 0 & 1 \end{pmatrix} \begin{pmatrix} \cos\gamma & 0 & \sin\gamma \\ 0 & 1 & 0 \\ -\sin\gamma & 0 & \cos\gamma \end{pmatrix} \begin{pmatrix} 0 \\ \sin\theta_{zy} \\ \cos\theta_{zy} \end{pmatrix}$$

$$= \begin{pmatrix} \cos\alpha \sin\gamma \cos\theta_{zy} - \sin\alpha \sin\theta_{zy} \\ \sin\alpha \sin\gamma \cos\theta_{zy} + \cos\alpha \sin\theta_{zy} \\ \cos\gamma \cos\theta_{zy} \end{pmatrix} \tag{6}$$

If the observed MR($\theta_{zy}$) is due to the misalignment of conventional AMR only, the angle dependent longitudinal resistivity should be given by:

$$\rho = \rho_0 + \Delta\rho(\cos\alpha \sin\gamma \cos\theta_{zy} - \sin\alpha \sin\theta_{zy})^2 \qquad (7)$$

where $\Delta\rho/\rho_0$ is the AMR ratio, about $2 \times 10^{-3}$ estimated from the ADMR results in Fig. 2 of the main text. Using Supplementary Equation 7, we calculated the contribution of AMR from field misalignment in the MR($\theta_{zy}$) with different combinations of misalignment angles up to 5°, and the results are plotted in Supplementary Figure 11. The solid line refers to the case where both rotations exist, and one of them is fixed at 5° with the other varying from 0° to 5°; while the dashed line is the case where only one of the rotation exists with an angle from 0° to 5°. As a comparison, the signal level of MR($\theta_{zy}$) of the FeMnPt sample observed in Fig. 2 of the main text is also added in the figure (dotted line). It should be noted that the misalignment angle above 5° is highly unlike in the present experimental setup. It is clearly that the size of the signal (about $10^{-5}$) is always nearly two orders of the magnitude smaller than that observed in Fe based samples. There must be another mechanism that gives rise to the MR($\theta_{zy}$) signal, that is, the AHMR.

In addition, to exclude GSE related AMR as the origin of MR($\theta_{zy}$) in Fe based samples, temperature dependent ADMR measurements were performed on (Fe$_{0.71}$Mn$_{0.29}$)$_{0.6}$Pt$_{0.4}$(9), Fe(9) and Py(9) samples. Despite its debatable underlying mechanism, GSE related AMR should still follow the temperature dependence of conventional AMR[17]. In other words, one should expect at least a same temperature dependence for MR($\theta_{zy}$) and MR($\theta_{zx}$) if MR($\theta_{zy}$) is dominated by GSE. The ADMR results of MR($\theta_{zy}$) and MR($\theta_{zx}$) in the temperature range of 50 – 300 K are presented in Supplementary Figures 12a – 12b for the FeMnPt sample, and the ratios are further summarized in Supplementary Figure 12c. In a similar manner, Supplementary Figures 12d – 12f present the results of the Fe sample, and Supplementary Figures 12i – 12h are those of the Py sample. In all three types of samples, we observed the increasing MR($\theta_{zx}$) ratio with the decrease of temperature, which supports the explanation that MR($\theta_{zx}$) is originated from the conventional AMR. This temperature dependence can be understood in the sense that the AMR ratio is enhanced by the reduction of phonon mediated $sd$ scattering at low temperature. As for the MR($\theta_{zy}$) ratio, we indeed observed the same increasing temperature dependence in the Py case. This supports our explanation that MR($\theta_{zy}$) in Py is dominated by GSE related AMR. In fact, similar observations have been reported and attributed to GSE in Co in the literature[17]. On the contrary, in Fe and Fe-based alloys, a totally opposite temperature dependence has been observed for MR($\theta_{zy}$) ratio, *i.e.*, it decreases with the decrease of temperature. This suggest that MR($\theta_{zy}$) and MR($\theta_{zx}$) in these samples have a different origin. As discussed in the main text, the AHMR ratio is given by $(\frac{\theta_{AH}}{\beta})^2 \frac{2l_s}{d} \tanh(\frac{d}{2l_s})$. As summarized in Supplementary Figure 13a for the FeMnPt sample, $\theta_{AH}$ can be obtained experimentally, and it is not very sensitive to temperature in the range of 50 – 300 K. On the other hand, although $\beta$ and $l_s$ is not directly accessible, one would expect both of them to increase with the decrease of temperature. To investigate their respective role on AHMR ratio, we calculated the $\beta$ and $l_s$ dependence of AHMR ratio in Supplementary Figure 13b ($\beta$ = 0.2 – 0.8, $l_s$ = 4.5 nm) and Supplementary Figure 13c ($\beta$ = 0.55, $l_s$ = 2 – 8 nm). In both figures, $\theta_{AH}$ is taken as 0.028, and the data in *x*- and *y*-axis are normalized to the minimum value in each axis. As can be seen, the increase of $\beta$ would lead to a decrease in AHMR ratio, whereas an opposite trend is obtained for $l_s$. However, for a same increase by a factor of 4, the effect of $\beta$ on AHMR is about 10 times larger than that of $l_s$. Therefore, the temperature dependence should be mainly determined by $\beta$, which agrees with the general trend of experimental temperature-dependence of AHMR in the FeMnPt samples. In fact, $\beta$ has also been found to play an important role in determining the temperature dependence of SMR in W/CoFeB bilayers[23]. Although further systematic studies are required to quantitatively elucidate the temperature dependence of AHMR, which is out of the scope of this manuscript, from the aforementioned experimental results and analysis, one can rule out GSE related AMR as the origin of MR($\theta_{zy}$).

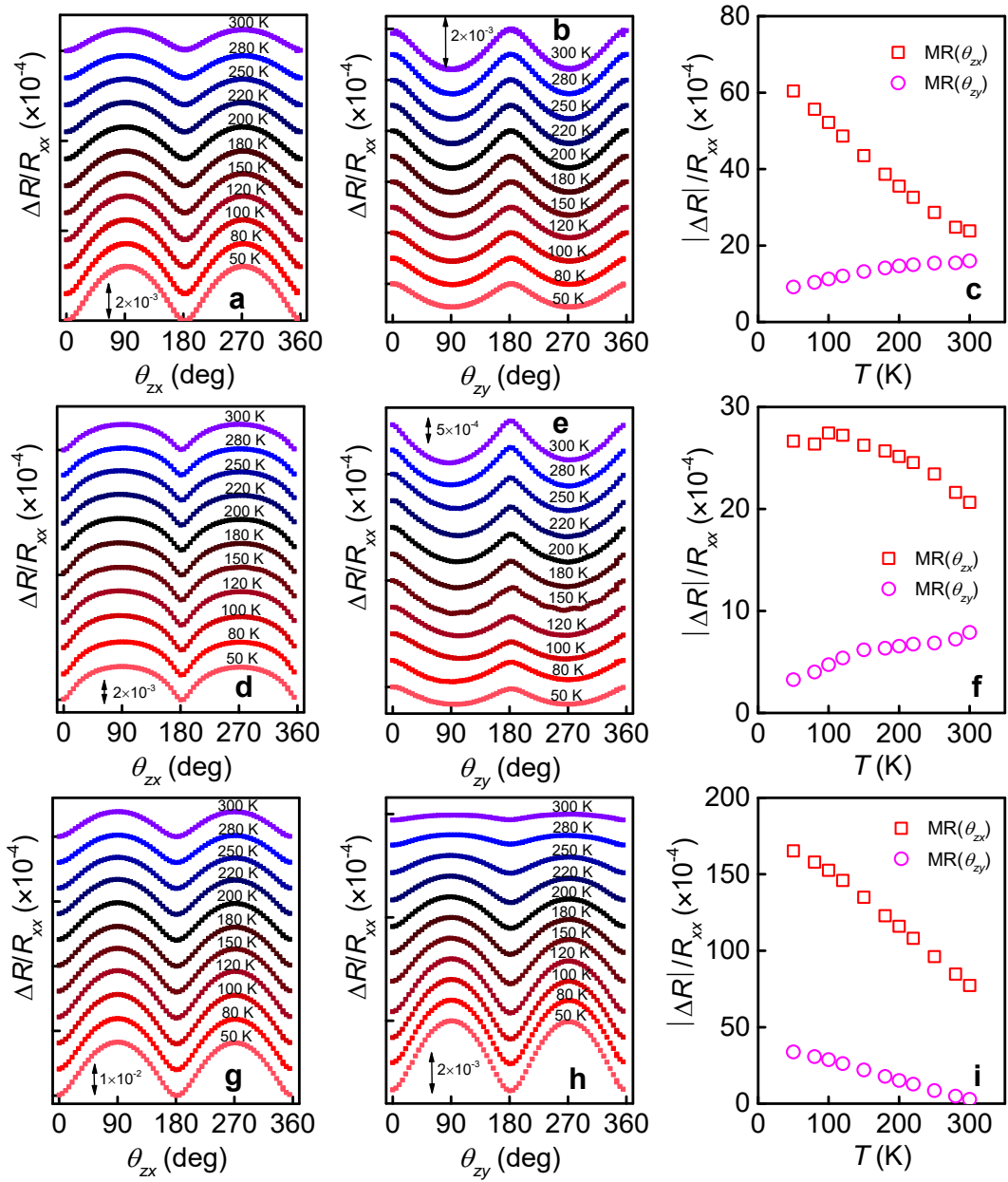

**Supplementary Figure 12. Temperature dependence of MR($\theta_{zy}$) and MR($\theta_{zx}$). a,** and **b,** ADMR results for (Fe$_{0.71}$Mn$_{0.29}$)$_{0.6}$Pt$_{0.4}$(9); **c,** Summary of the MR ratios for (Fe$_{0.71}$Mn$_{0.29}$)$_{0.6}$Pt$_{0.4}$(9); **d,** and **e,** ADMR results for Fe(9); **f,** Summary of the MR ratios for Fe(9); **g,** and **h,** ADMR results for Py(9); **i,** Summary of the MR ratios for Py(9).

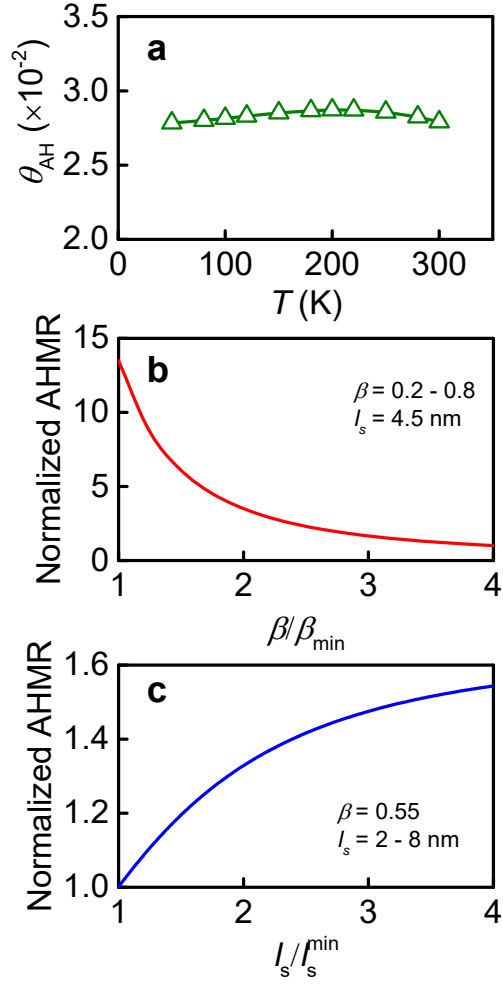

**Supplementary Figure 13. Effect of $\theta_{AH}$, $\beta$ and $l_s$ on the temperature dependence of AHMR. a,** Experimentally obtained temperature dependence of $\theta_{AH}$ in $(Fe_{0.71}Mn_{0.29})_{0.6}Pt_{0.4}(9)$; **b,** Calculated $\beta$ dependence of AHMR ratio with $\beta = 0.2 – 0.8$, $l_s = 4.5$ nm; **c,** Calculated $l_s$ dependence of AHMR ratio with $\beta = 0.55$, $l_s = 2 – 8$ nm.

**Supplementary Note 8. Derivation of anomalous Hall magnetoresistance (AHMR)**

To model the AHMR effect, we begin with the definition of the spin Hall angle $\theta_{SH}$ and the anomalous conductivity $\sigma_{xy}^{AH}$ for a bulk conducting FM:

$$j_{\uparrow}^{t} = \theta_{SH} j_{\uparrow} \tag{8}$$

$$j_{\downarrow}^{t} = -\theta_{SH} j_{\downarrow} \tag{9}$$

where $j^t_{\uparrow,\downarrow}$ are the transverse current induced by the longitudinal current $j_{\uparrow,\downarrow}$. The anomalous transverse charge ($j^t_c$) and spin current ($j^t_s$) is then,

$$j^t_c = j^t_\uparrow + j^t_\downarrow = \theta_{SH}\beta j_c \tag{10}$$

$$j^t_s = j^t_\uparrow - j^t_\downarrow = \theta_{SH}(j_\uparrow + j_\downarrow) = \theta_{SH} j_c \tag{11}$$

where $\beta = (j_\uparrow - j_\downarrow)/(j_\uparrow + j_\downarrow)$ is the spin polarization of the longitudinal current. The non-zero charge current would lead to a charge buildup at the side of the conductance which would exactly cancel the transverse charge current. Thus, the ratio of the anomalous conductivity $\sigma^{AH}_{xy}$ to the longitudinal conductivity $\sigma_{xx}$ (or anomalous Hall angle) is $\theta_{AH} \equiv \sigma^{AH}_{xy}/\sigma_{xx} \equiv j^t_c/j_c = \beta\theta_{SH}$. It should be noted that from Supplementary Equation 10 and 11, the spin polarization of the transverse conductivity ($\zeta$) can be inferred as $\zeta = j^t_s/j^t_c = 1/\beta$. In practice, $\zeta$ may not be related to $\beta$ in this simple way[24, 25]: $\beta$ is determined by the spin-dependent density of states at Fermi level and scattering asymmetry between spin-up and spin-down electrons in FM and its value is always between -1 and 1; while $\zeta$ is not entirely determined by the same mechanisms since the spin-up and spin-down electrons may not be deflected to opposite directions with equal probability due to the fact that scattering potentials seen by the two types of electron are different. However, as discussed by Taniguchi et al.[26, 27], if one ignores the energy-dependence of scattering and assume that the deflected spin-up and spin-down electrons will undergo the same spin-dependent scattering as the longitudinal transport, then $\zeta$ can be related to $\beta$ as $\zeta = 1/\beta$. This treatment largely simplifies the relation between $\theta_{AH}$ and $\theta_{SH}$ as is shown above.

Next we turn to look at how the spin current can affect the longitudinal resistance. The transverse spin current $j^t_s$ freely flows in the bulk, but would lead to spin accumulation at the edge of the sample. Different from the charge accumulation which acts on the entire sample to cancel the transverse current completely, the spin accumulation only acts on the vicinity of the sample boundary and thus the spin current in the interior of the sample remains to be $j_c \theta_{AH} / \beta$. The transverse spin current in turn can generate a charge current whose flow direction is always opposite to the original charge current. Thus the total charge current would be $j_c - j_c(\theta_{AH}/\beta)^2$, and it increases the resistance of the film by a factor of $[1-(\theta_{AH}/\beta)^2]^{-1} = 1+(\theta_{AH}/\beta)^2$ due to the combined charge to spin and spin to charge conversion.

Now we apply the above procedure to a thin FM film with thickness of $d$ in $z$-direction, and the applied current is fixed at $x$-direction. With the presence of large exchange field, the spins in FM are either parallel or antiparallel to the magnetization direction ($\mathbf{m}$). Bear this in mind, below we further discuss the three cases with $\mathbf{m}$ aligned in different directions by external magnetic field.

When $\mathbf{m}$ is in $z$-direction, the spin current flows in $y$-direction and the spin accumulation is at the front and back edges of the film. Due to the much larger dimension of the film width as compare to the spin diffusion length ($l_s$), the spin current is not affected inside the film and thus the resistance would be the same as the bulk FM case. Therefore, the total transverse spin current, longitudinal charge current and resistivity are summarized respectively as

$$j_s^t(\mathbf{m} \parallel \mathbf{z}) = j_c \theta_{AH} / \beta \tag{12}$$

$$j_{cx}(\mathbf{m} \parallel \mathbf{z}) = j_c - j_c(\theta_{AH}/\beta)^2 \tag{13}$$

$$\rho_{xx}(\mathbf{m} \parallel \mathbf{z}) = \rho_0[1+(\theta_{AH}/\beta)^2] \tag{14}$$

where $\rho_0$ is the isotropic resistivity of FM.

When $\mathbf{m}$ is in $y$-direction, the spin current flows in $z$-direction. In this case, the spin is accumulated at the surface or interface of the film. Since now $d$ is comparable to $l_s$, the spin accumulation leads to a backflow of spin current which would greatly reduce the total spin current, and it cancels some of the AHE induced extra resistance. The general solution of spin diffusion equation ($\partial_z^2 \mu_s = \mu_s / l_s^2$) is $\mu_s(z) = A e^{z/l_s} + B e^{-z/l_s}$. By using the boundary conditions: $j_s^t(0) = j_s^t(d) = 0$, we can derive the spin accumulation and transverse spin current in $z$-direction as:

$$\mu_s(z) = \frac{2 e l_s j_c}{\sigma_{xx}} \frac{\theta_{AH}}{\beta} \left( \cosh\left(\frac{z}{l_s}\right) - \cosh\left(\frac{z-d}{l_s}\right) \right) / \sinh\left(\frac{d}{l_s}\right) \quad (15)$$

$$j_s^t(z) = \frac{j_c \theta_{AH}}{\beta} \left[ 1 - \left( \sinh\left(\frac{z}{l_s}\right) - \sinh\left(\frac{z-d}{l_s}\right) \right) / \sinh\left(\frac{d}{l_s}\right) \right] \quad (16)$$

$$j_{cx}(z) = j_c - j_c \left(\frac{\theta_{AH}}{\beta}\right)^2 \left[ 1 - \left( \sinh\left(\frac{z}{l_s}\right) - \sinh\left(\frac{z-d}{l_s}\right) \right) / \sinh\left(\frac{d}{l_s}\right) \right] \quad (17)$$

Average the above spin current over the thickness, one has the total transverse spin current, longitudinal charge current and resistivity in this case as

$$j_s^t(\mathbf{m} \parallel \mathbf{y}) = j_c (\theta_{AH}/\beta) \left( 1 - \frac{2 l_s}{d} \tanh(d/2 l_s) \right) \quad (18)$$

$$j_{cx}(\mathbf{m} \parallel \mathbf{y}) = j_c - j_c (\theta_{AH}/\beta)^2 \left( 1 - \frac{2 l_s}{d} \tanh(d/2 l_s) \right) \quad (19)$$

$$\rho_{xx}(\mathbf{m} \parallel \mathbf{y}) = \rho_0 \left[ 1 + (\theta_{AH}/\beta)^2 \left( 1 - \frac{2 l_s}{d} \tanh(d/2 l_s) \right) \right] \quad (20)$$

When $\mathbf{m}$ is in $x$-direction, there is no transverse spin current and thus no extra-resistance. However, the conventional anisotropy magnetoresistance (AMR) would appear, and this leads to the total transverse spin current, longitudinal charge current and resistivity as $j_s^t(\mathbf{m} \parallel \mathbf{x}) = 0$, $j_{cx}(\mathbf{m} \parallel \mathbf{x}) = j_c$ and $\rho_{xx}(\mathbf{m} \parallel \mathbf{x}) = \rho_0 (1 + A)$, where $A$ is the AMR ratio. Taken together the above three cases, one can summarize the MR effect in a single FM as

$$\rho_{xx} = \rho_0 \left( 1 + A m_x^2 + (\theta_{AH}/\beta)^2 \left[ m_z^2 + \left( 1 - \frac{2 l_s}{d} \tanh(d/2 l_s) \right) m_y^2 \right] \right) \quad (21)$$

**Supplementary Note 9. Effect of surface roughness and film thickness on transport and magnetic properties in ultrathin films**

It is known that the percolated structure or significant surface roughness in very thin films can affect both the electrical and magnetic properties. In addition, any change in the surface condition after the sample was exposed to ambient may also affect its physical properties. As the films under investigation are polycrystalline in nature, it would be difficult to achieve layer-by-layer growth at atomic layer accuracy and therefore, the presence of a certain degree of roughness is unavoidable. We have previously investigated systematically the electrical properties of ultrathin metallic film[28], including Al, Au, Cr, Cu, Ru, Ta, $Co_{90}Fe_{10}$, $Ni_{81}Fe_{19}$, and $Ir_{20}Mn_{80}$. Different materials indeed exhibit different level of roughness. Except for Al, Au and Cu, the root-mean-square (RMS) roughness of remaining films at a thickness of 20 nm is generally below 0.2 nm. The resistivity of all these films show an upturn at small thickness, though the turning point is generally more than one order of magnitude larger than the roughness. This suggests that the thickness at which sharp upturn of resistivity appears is mainly determined by the electron mean free path, as we discussed in main text.

To characterize the roughness of the thin film used in this study, we preformed atomic force microscopy (AFM) measurement on the 5 nm sample of $(Fe_{0.71}Mn_{0.29})_{0.6}Pt_{0.4}$ and Fe. As an example, Supplementary Figure 14 shows the AFM image for $(Fe_{0.71}Mn_{0.29})_{0.6}Pt_{0.4}$ within an area of 5 μm × 5 μm. The averaged RMS roughness over 5 different areas with such a size is 0.26 nm and 0.33 nm for $(Fe_{0.71}Mn_{0.29})_{0.6}Pt_{0.4}$ and Fe, respectively. As shown in Fig. 5b of the main text, the sharp upturn of resistivity in $(Fe_{0.71}Mn_{0.29})_{0.6}Pt_{0.4}$ appears at about 3 nm, which is also about 10 times of the RMS roughness, in good agreement with previous studies. Therefore, the sharp upturn of resistivity kicks in when the thickness of the film becomes comparable to the electron mean free path rather than due to reaching the percolation threshold of forming discontinuous film. In fact, the resistivity values of $(Fe_{0.71}Mn_{0.29})_{0.6}Pt_{0.4}$ and Fe with a thickness of 5 – 20 nm are in the range of $3.9×10^3 – 1.8×10^4$ $(\Omega\ cm)^{-1}$ and $4.9\times10^3 - 2.3\times10^4$ $(\Omega\text{ cm})^{-1}$, respectively. These values fall into the upper bound of bad metal and lower bound of good metal regime[14]. Therefore, we can say that both the $(Fe_{0.71}Mn_{0.29})_{0.6}Pt_{0.4}$ and Fe films with a thickness of 5 – 20 nm are continuous metallic films.

On the other hand, both surface and size-effect also affect magnetic properties of thin films, which typically would lead to decrease of saturation magnetization. There is no generic model to describe the thickness dependence of saturation magnetization in ultrathin films since both the surface and interface with substrate vary from sample to sample. As far as $(Fe_{0.71}Mn_{0.29})_{0.6}Pt_{0.4}$ thin film is concerned, as shown in Fig. 5b of the main text, the magnetization began to decrease at a thickness of ~ 5 nm, which is larger than the thickness at which the resistivity shows a sharp upturn. This is understandable since they are governed by phenomena of different length scale. However, as explained in the main text, this does not affect the analysis and interpretation of the experimental data of films with $d >$ 5 nm. For samples with $d <$ 5 nm, the experimental data can be understood qualitatively if we take into account the thickness-dependent $\theta_{AH}$ obtained experimentally. However, we did not include the fitting results in Fig. 5a of the main text because thickness-dependence of $\beta$ is unknown both theoretically and experimentally.

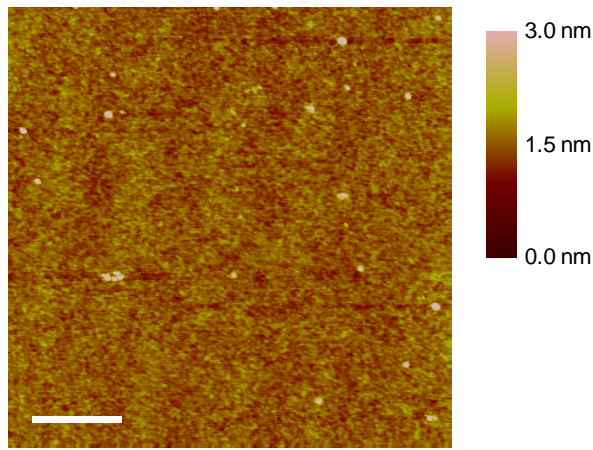

**Supplementary Figure 14. Characterization of thin film roughness.** AFM image of a 5 nm-thick $(Fe_{0.71}Mn_{0.29})_{0.6}Pt_{0.4}$. The scale bar is 1 μm.

**Supplementary Note 10. Thickness dependence of AHMR in thick $(Fe_{0.71}Mn_{0.29})_{0.6}Pt_{0.4}$ and Fe samples**

To substantiate the thickness dependence of AHMR, besides the results shown in Fig. 5 of the main text, we fabricated another batch of $(Fe_{0.71}Mn_{0.29})_{0.6}Pt_{0.4}$ and extended the same thickness dependence study to Fe as well. Supplementary Figures 14a and 14b show the MR($\theta_{zy}$) curves for $(Fe_{0.71}Mn_{0.29})_{0.6}Pt_{0.4}$ and Fe samples. To avoid any ambiguity, this time we focused on the thickness range of 5 – 20 nm, where the ferromagnetic and electrical properties of the films is almost unchanged. This is also evident in the plot of $\rho_{xy}^{AH} / M_s$ as a function of $\rho_{xx}$ in Supplementary Figure 14c, which exhibits an almost linear scaling. As summarized in Supplementary Figure 14d, despite some variations in the absolute values at some thicknesses (due to some slight differences in measurement environment for different runs), the general trend of MR($\theta_{zy}$) is the same as that presented in Fig. 5, *i.e.* AHMR decreases as film thickness increases for $d > 5$ nm. This is in good agreement with the thickness dependence predicted by AHMR theory. In fact, by taking $\beta = 0.62$, $l_s = 6.5$ nm, $\theta_{AH} = 0.027$ for FeMnPt, and $\beta = 0.32$, $l_s = 3.2$ nm, $\theta_{AH} = 0.009$ for Fe, both sets of data can be fitted well to the theoretical model. It should be noted that the differences in the parameters used here and those in the main text for FeMnPt may be caused by the differences in the detailed sample preparation and measurement processes, as well as surface conditions. Nevertheless, in view of all these thickness dependence results, one can see that MR($\theta_{zy}$) does follow the theoretical model.

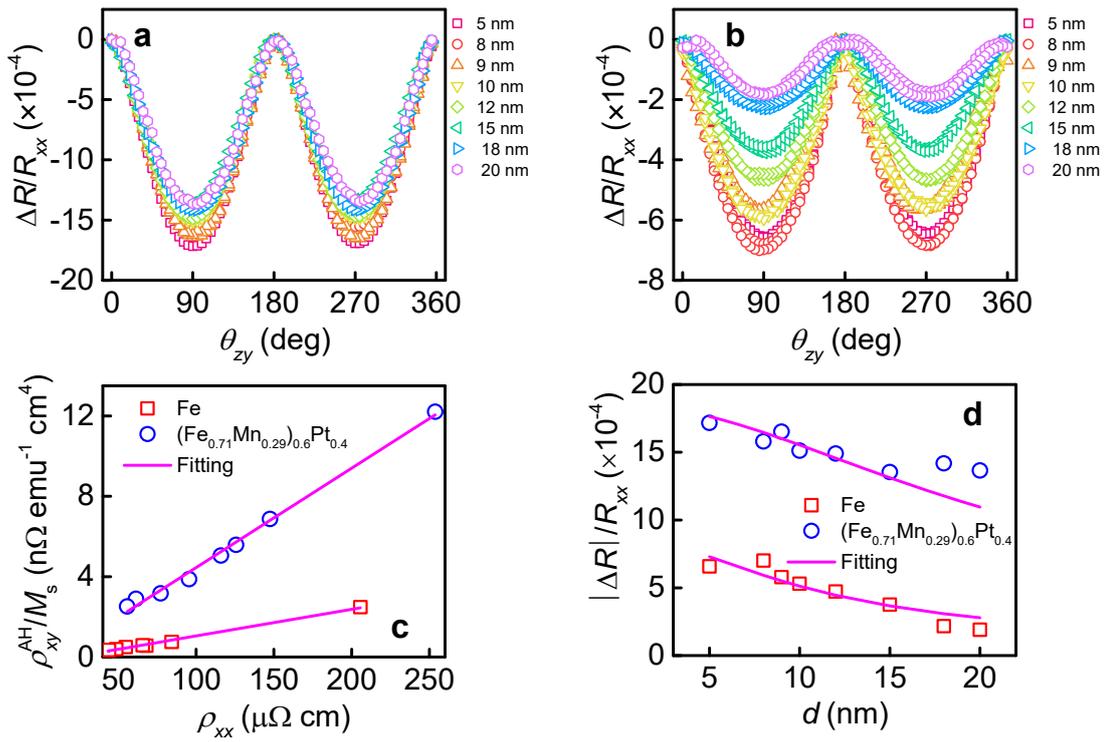

**Supplementary Figure 15. Thickness dependence of AHMR in the thick thickness region. a,** and **b,** MR($\theta_{zy}$) results for the new batch of (Fe$_{0.71}$Mn$_{0.29}$)$_{0.6}$Pt$_{0.4}$ and Fe, respectively; **c,** Plot of $\rho_{xy}^{AH}/M_s$ as a function of $\rho_{xx}$ in the (Fe$_{0.71}$Mn$_{0.29}$)$_{0.6}$Pt$_{0.4}$ and Fe samples; **d,** Summary of the thickness dependence of MR($\theta_{zy}$) and fitting using the theoretical model.